\documentclass[fleqn,12pt]{wlscirep}
\usepackage[utf8]{inputenc}
\usepackage[T1]{fontenc}

\usepackage{multirow}
\usepackage{subcaption}
\usepackage{graphicx}
\usepackage{amssymb}
\usepackage{amsmath}
\usepackage{wasysym}
\usepackage{rotating}

\usepackage[normalem]{ulem}

\title{The arrangement of anisotropic spin couplings can optimize sensitivity of 
the cryptochrome radical pair to the direction of geomagnetic field}

\author[1,*]{Victor Bezchastnov}

\author[2]{Tatiana Domratcheva}

\affil[1]{Max Planck Institute for Medical Research, 
Department of Biomolecular  Mechanisms, Jahnstrasse 29, 69120 Heidelberg, Germany}

\affil[2]{Lomonosov Moscow State University, Department of Chemistry, 
119991 Moscow, Russia}

\affil[*]{victor.beschastnov@mpimf-heidelberg.mpg.de}

\begin{abstract}
Sensing of the geomagnetic field direction by many living organisms is commonly 
thought to involve radical pairs, such as those formed photochemically between the 
flavin and tryptophan radicals in the cryptochrome proteins. 
Previous theoretical studies have shown that strongly axial hyperfine couplings 
in the cryptochrome radicals greatly enhance the formation of a signaling state of 
the protein when the magnetic field is directed perpendicular to the hyperfine axis 
of either of the radicals. However, further analysis led to the conclusion that sharpness 
of detecting those magnetic directions is strongly suppressed by the inter-radical 
electron spin coupling. Here, we perform theoretical simulations of the compass function 
for a set of arrangements of the intra- and inter-radical spin couplings in the 
idealized cryptochrome radical pair, and find certain arrangements that 
preserve the sharpness in detecting the direction of the geomagnetic field. One 
particular arrangement, with the hyperfine axes of the radicals orthogonal to the 
symmetry axis of inter-radical coupling, provides even sharper field-direction 
sensitivity than that contributed solely by the anisotropy of the hyperfine coupling.
\end{abstract}

\begin{document}
\flushbottom
\maketitle

\section{Introduction}
\label{Intro}

The ability of many living organisms to sense the Earth’s magnetic field is thought to 
rely on magneto-sensitive photochemical reactions in cryptochrome photoreceptor proteins: 
see, e.g., the review~\cite{Hore-2016-07-05} and the references therein. In particular, 
it is believed that cryptochrome photoreceptors located in the 
retina~\cite{Ritz-2000-02-01,Mouritsen-2004-09-28,Moeller-2004-12-01,Zapka-2009-10} 
enable visual perception of the direction of the geomagnetic poles~\cite{Wiltschko-2009-10-28} 
by migratory birds. The underlying mechanism involves a transient radical pair which is formed 
by photoactivation of a receptor protein and whose spin dynamics mediate subsequent 
formation of a signaling state sensitive to the magnetic field 
direction~\cite{Ritz-2000-02-01,Rodgers-2009-01-13}. 
The study~\cite{Rodgers-2009-01-13} supposed that a strongly anisotropic but relatively 
simple direction information would be favored for achieving the optimum sensory effect. 

In the cryptochrome, the radical pair is formed between a bound flavin adenine 
dinucleotide (FAD) chromophore and one of the three or four tryptophan (Trp) residues 
that bridge the chromophore to the surface of the protein. The spin character of the 
two unpaired electrons, initially the same as of the photo-exited singlet FAD chromophore, 
varies during the lifetime of the pair, whereas the spin-selective decay of the radical 
pair determines the formation of the signaling state. As the electron spins are coupled 
to the magnetic field, the decay yield of the signaling state becomes dependent on the 
magnetic field vector. A key part of the magneto-sensitivity supported in this way is the 
coherent behavior of the unpaired electron spins and the associated interconversion of 
singlet and triplet states affected by the magnetic field. A theoretical description 
of the underlying physics and an accurate simulation of anisotropic magnetic field effects 
seem to require ultimately a full quantum-mechanical framework~\cite{Fay-2019-12-13}.

The radical pair mechanism of magnetoreceptions operates as an inclination compass by 
producing a response according to the angle defining the orientation of the radical pair 
in the magnetic field. The mechanism does not distinguish between two opposite directions 
of the magnetic field, in particular the directions to the south and north magnetic poles. 
The same type of magnetic orientation, which is rather of the inclination than of the 
polarity type, was shown by migratory birds in the behavioral 
experiments~\cite{Wiltschko-1972-04-07,Wiltschko-2001-10-01,Wiltschko-2006}. 
In theoretical simulations it remains difficult to replicate the high precision of 
the radical pair compass that is evident from the 
experimental finding~\cite{Akesson-2001-09-22,Lefeldt-2015-01-15} 
that the angular uncertainty of detecting the geomagnetic field lines by birds 
does not exceed $5^\circ$.

The spin dynamics for a magnetically sensitive photoreceptor have been the subject of 
many theoretical studies, e.g. 
\cite{Rodgers-2009-01-13,Fay-2019-12-13,
Timmel-1998-09-01,
Timmel-2001-02-09,
Solovyov-2007-04-15,
Lee-2014-06-06,
Hiscock-2016-04-26,
Kattnig-2016-05-04,
Pedersen-2016-11-10,
Atkins-2019-08-09,
Procopio-2020-02-11,
Babcock-2020-04-02,
Bezchastnov-2023-01-20}, 
addressing the putative radical pairs with different structures and spin interactions. 
The relevant interactions are modelled in terms of the {\em intra-radical} hyperfine 
coupling (HFC) and the {\em inter-radical} electron-electron dipolar (EED) and exchange 
couplings. The HFC results from the interaction of the spin of an unpaired electron 
in each radical with the spins of the nuclei possessing the spin magnetic moments, 
and the EED and exchange couplings account for the interaction of the unpaired 
electronic spins of two radicals with each other. The spin interaction in the 
magnetoreceptor necessarily involves the Zeeman coupling of the unpaired electronic 
spins to the magnetic field.

Couplings of the different types play distinct roles in the spin dynamics. 
The HFC is crucial for inducing the spin dynamics as 
such~\cite{Brocklehurst-1974-10-01,Schulten-1976-09-01,Schulten-1978-01-01}, and the 
anisotropy of the HFC is an important prerequisite for the influence of the field 
direction on the dynamics and their final yield~\cite{Schulten-1978-01-01,Fay-2019-12-13}. 
In particular, the HFC of the cryptochrome flavin-tryptophan radical pair contains highly 
axial contributions from the nuclei of the nitrogen atoms N5, N10 of FAD, and N1 of Trp, 
which can lead to a sharp effect of the field direction on the spin dynamics and the 
yield of the spin-selective decay. The study~\cite{Hiscock-2016-04-26} demonstrated 
such an effect, referred to as a ``quantum needle'', as sharp, spiky features in the 
dependence of the decay yield on the direction of the $50~\mu$T magnetic field. 
The simulations performed for the radical pair with 16 nuclear spins revealed that 
the most intense spike arises for the field directions in the plane of the flavin 
indole rings and reflects the part of the flavin HFC approximately symmetric around 
the axis orthogonal to the rings. A smaller spike was found to appear for the magnetic 
field in the plane of the indole ring of the tryptophan, reflecting the highly axial 
portion of the tryptophan HFC. Similar simulations, such as 
\cite{Pedersen-2016-11-10,Babcock-2020-04-02,Worster-2016-07-20,Hiscock-2017-10-03}, 
described spikes of the same nature, and the study~\cite{Bezchastnov-2023-01-20} 
provided a quantum-mechanical insight into the spin-state properties responsible for 
the spikes. The two distinct spike features become more pronounced and equal in 
intensity for the model radical pair with HFC exerted only by the FAD N5 and Trp N1 
nuclei. Analyzing the latter model, it was pointed out~\cite{Bezchastnov-2023-01-20} 
that, while the spikes correspond to the different manifolds of the magnetic field 
directions related to the symmetries of FAD N5 and Trp N1 HFC contributions, there are 
two opposite directions of the field, for which the two spikes merge into one. This 
naturally defines a {\em particular direction} of the magnetic field relative to the 
radical pair, the detection of which might be used to obtain precise compass bearings 
for the directions of the magnetic poles.

The quantum needle effect is an important finding since narrow spikes can be considered 
as the basis of the high accuracy of the radical-pair based compass. However, 
this effect, being related to the intra-radical HFC anisotropy and first 
observed~\cite{Hiscock-2016-04-26} when simulating the spin dynamics in the absence of 
the inter-radical spin interaction, turned out to significantly deteriorate when the 
latter interaction was taken into account. For the cryptochrome radical pairs, the 
experimental studies~\cite{Nohr-2017,Xu-2021-06} indicated a small exchange 
but a significant EED spin coupling, and the theoretical 
simulations~\cite{Babcock-2020-04-02} demonstrated that the latter strongly 
suppresses the HFC-related spikes. Before the 
studies~\cite{Babcock-2020-04-02,Nohr-2017,Xu-2021-06}, 
the analysis~\cite{Efimova-2008-03-01} of the inter-radical interaction in the model of 
the photoreceptor compass demonstrated that the exchange and EED spin interactions 
can partially compensate for each other’s detrimental influence on compass sensitivity. 
With such compensation and with the HFC simplified to the 
contribution of a single proton in one of the radicals, the dependence of decay yield 
on the magnetic field direction was notably anisotropic, reflecting 
the symmetry of the HFC as well as of the EED coupling. However, for a more realistic 
model of the cryptochrome radical pair with highly axial HFC contributions from the 
nitrogen nuclei FAD N5, N10, and Trp N1, the simulations~\cite{Babcock-2020-04-02} 
demonstrated that the mutual exchange/EED compensation is inefficient not only in 
re-establishing the spikes but also in supporting an appreciable anisotropy of the 
response to the magnetic field direction. 
In addition, the spin interactions are affected by thermal motions, 
resulting in the electron spin relaxation, as described in a theoretical 
study~\cite{Lau-2009-12-09}. The latter dynamical effects generally 
weaken the compass function~\cite{Kattnig-2016-05-04,Worster-2016-07-20}, but, under 
certain conditions, can also enhance the performance of the cryptochrome radical-pair 
magnetic sensor~\cite{Kattnig-2016-06-09,Benjamin-2025-06-04}. 

In the present work we studied how the arrangement of the spin interactions affects 
the response of a model cryptochrome radical pair with spatially fixed atomic nuclei 
to the direction of a magnetic field. We aimed at accurate 3D visualisation of the 
narrow spikes in the direction-dependence of the response. 
Based on the description~\cite{Hore-2016-07-05} of the radical-pair 
mechanism of the magnetoreception, we consider the singlet-born 
flavin-tryptophan radical pair and, as in the previous 
study~\cite{Bezchastnov-2023-01-20}, focus on sensitivity of the 
singlet-triplet interconversion, i.e., of the core part of the mechanism, to the 
direction of the geomagnetic field with a typical magnitude of $50~\mu$T. We also 
address a decaying radical pair by computing the yield of the signaling state for a 
particular choice of the decay rate constants and field directions. 
We consider first the radical pair with the entire spin interaction restricted to 
the highly axial FAD N5 and Trp N1 HFC contributions, which produce prominent spiky 
features in the anisotropy of the magnetic response. Expanding the spin interaction in 
this radical pair by the EED spin coupling entirely suppresses the response anisotropy, 
in spite of the axial shapes of both the hyperfine and EED couplings. 
However, adding the electron spin exchange interaction 
of a particular~\cite{Efimova-2008-03-01} strength restores the response 
anisotropy but not the spikes, with the anisotropy pattern indicating the direction 
along the EED coupling. Next, we find that an orientation of the hyperfine couplings 
orthogonal to the EED coupling sharpens the anisotropy of the response by the spikes, 
making such a radical pair even superior, as compared to that without the inter-radical 
spin coupling, in precise detection of the field direction. Finally, we demonstrate 
that expanding HFC preserves the spiky part of the response anisotropy, as long as 
the additional HFC contributions do not destroy the overall highly axial anisotropy 
of the hyperfine interaction in each radical.

\section{Radical pair models and theoretical framework of quantum simulations}
\label{Theory}

The details of theoretical simulations are described below for basic models of 
the $\left[\mbox{FAD}^{\bullet{-}}-\mbox{Trp}^{\bullet{+}}\right]$ radical pair, 
in which the intra-radical spin coupling is determined by the 
Zeeman coupling of the electron spins to the magnetic field, and by the hyperfine 
coupling of the electron spin to the nucleus spin of a single nitrogen atom in each 
radical: N5 in FAD$^{\bullet{-}}$ and N1 in Trp$^{\bullet{+}}$. Applying such an 
approach to radical pairs with expanded HFC contributions is straightforward. 
We consider four basic models, referred to as RP1, RP2, RP3, and RP4: 
RP1 contains the intra-radical coupling only; RP2, RP3 and RP4 include the 
inter-radical EED coupling; and RP3 and RP4 additionally take account of the 
coupling due to the electron exchange. The intra- and inter-radical spin 
interactions constitute the Hamiltonian
\begin{equation}
H = A_{ij}^{({\rm N5})}s_i^{(1)}L_j^{({\rm N5})}
  + A_{ij}^{({\rm N1})}s_i^{(2)}L_j^{({\rm N1})}
  + B_i \left( s_i^{(1)} + s_i^{(2)} \right) 
  + D_{ij}s_i^{(1)}s_j^{(2)}
  + J s_i^{(1)}s_i^{(2)},
\label{H}
\end{equation}
in which $A_{ij}^{({\rm N5})}$ and $A_{ij}^{({\rm N1})}$ are the flavin N5 and 
tryptophan N1 HFC tensors, $L_j^{({\rm N5})}$ and $L_j^{({\rm N1})}$ are the components 
of the spins of the two nitrogen nuclei, $D_{ij}$ is the EED coupling tensor, 
$J$ is the electron spin exchange constant, $s_i^{(1)}$ and $s_i^{(2)}$ are the 
components of the spin of the unpaired electron of the flavin (1) and tryptophan (2), 
and $B_i$ are the components of the magnetic field. The indices $i,j$ designate the 
axes of a coordinate reference frame, and the tensor notations assume summations over 
the repeating indices. We use the units of mT for the $A_{ij}^{({\rm N5})}$, 
$A_{ij}^{({\rm N1})}$, $D_{ij}$ and $J$ ($1$~mT corresponds to $4.253\times{10}^{-9}$ 
hartree), and neglect a small deviation of the electron $g$-factor from $2$ in the 
Zeeman coupling. The response of the radical pair to direction of the magnetic field 
is simulated for the field magnitude of $50~\mu$T.

In our models we use the spin couplings derived for the radical pair formed by the 
flavin and the tryptophan that is third-closest to it in the 
structure~\cite{Levy-2013-03} of the {\em Drosophila} cryptochrome protein. 
The couplings are determined by the densities of the unpaired electrons that we 
computed taking into account the structural changes in the atomic clusters due to the 
electron transfer from the tryptophan to the flavin. The couplings obtained thus 
correspond to the atomic locations in the FAD$^{\bullet{-}}$ and Trp$^{\bullet{+}}$ 
radicals, which are slightly different from the respective locations in the unexcited 
cryptochrome structure. The computed FAD N5 and Trp N1 HFC tensors agree well with 
the tensors reported in other studies, e.g.~\cite{Hiscock-2016-04-26,Deviers-2022-07-13}. 
The two tensors have similar anisotropies, described by the relation between their 
three principal values, $A_1 \approx A_2$, $\left| A_1 \right| \ll A_3$, so that the 
tensor shapes are nearly axially symmetric and elongated in the principal 
direction ${\bf e}_3$, which we refer to as the {\em HFC axis}. For each radical, 
the orientation of the HFC axis is very close to the orthogonal to the heterocyclic 
plane one (see Figure~\ref{fig_1}): the angle between the N5 axis and the 
FAD$^{\bullet{-}}$ plane is $86.9^\circ$, the angle between the N1 axis and the 
Trp$^{\bullet{+}}$ plane is $84.3^\circ$, and the angle between the axes is $31^\circ$. 
To derive the EED spin coupling tensor, we employ the point-dipole approximation 
justified by a large separation of the radicals. The tensor is determined by the 
distance between the centers of the densities of two unpaired electronic spins, and 
is symmetric around the {\em EED axis} defined by the principal direction ${\bf e}_1$, 
which points from one of the centers to the other. The separation distance 
$16.305$~{\AA} agrees with the radical separation determined for the {\em Drosophila} 
cryptochrome by EPR spectroscopy~\cite{Nohr-2017}. In this study, we examine 
a particular case, where the EED and exchange couplings partially cancel each other 
in affecting the radical-pair spin states. To this end, we take the $J$ value equal to 
the negative of the $D_2 = D_3$ principal value of the EED spin coupling tensor: this 
choice yields degenerate S and T$_0$ states of the system of two electron 
spins~\cite{Efimova-2008-03-01}.
%
% Figure 1
%
\bigskip
{\linespread{1.2}
\begin{figure}[h]
\centering
\begin{subfigure}{0.24\textwidth}
    \includegraphics[width=\textwidth]{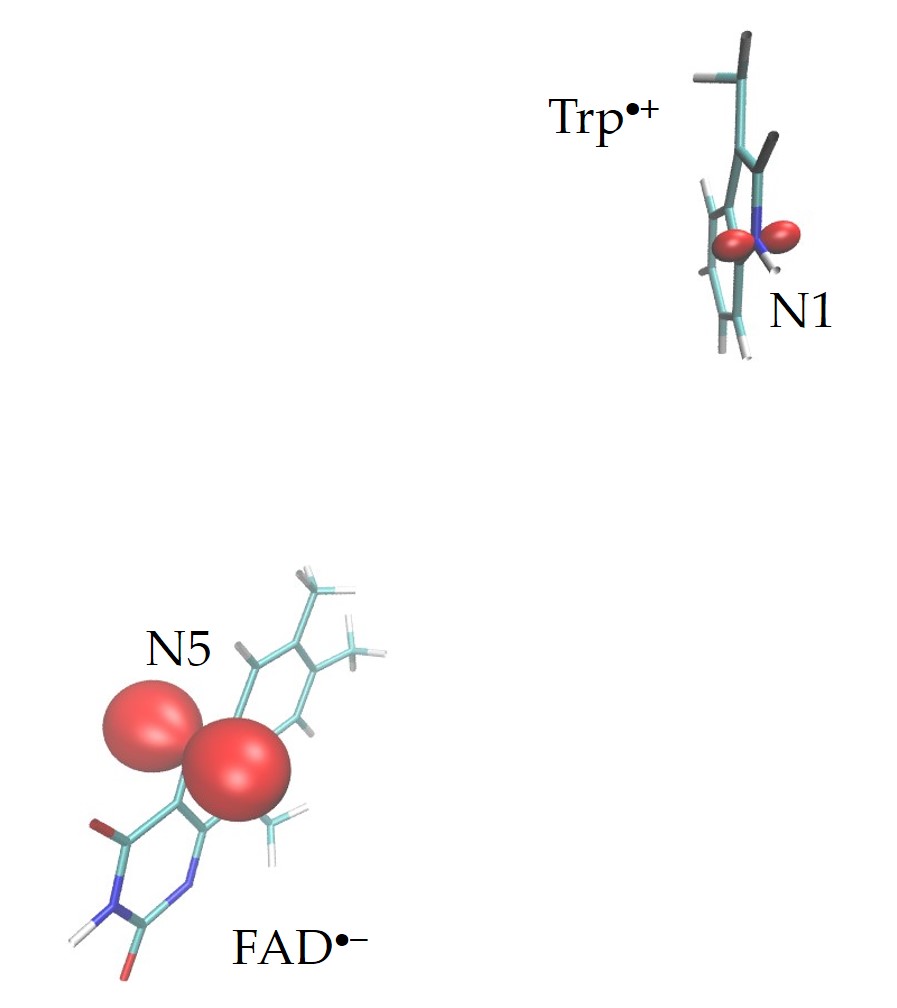}
    \caption*{(1)}
    \label{RP1}
\end{subfigure}
\hfill
\begin{subfigure}{0.24\textwidth}
    \includegraphics[width=\textwidth]{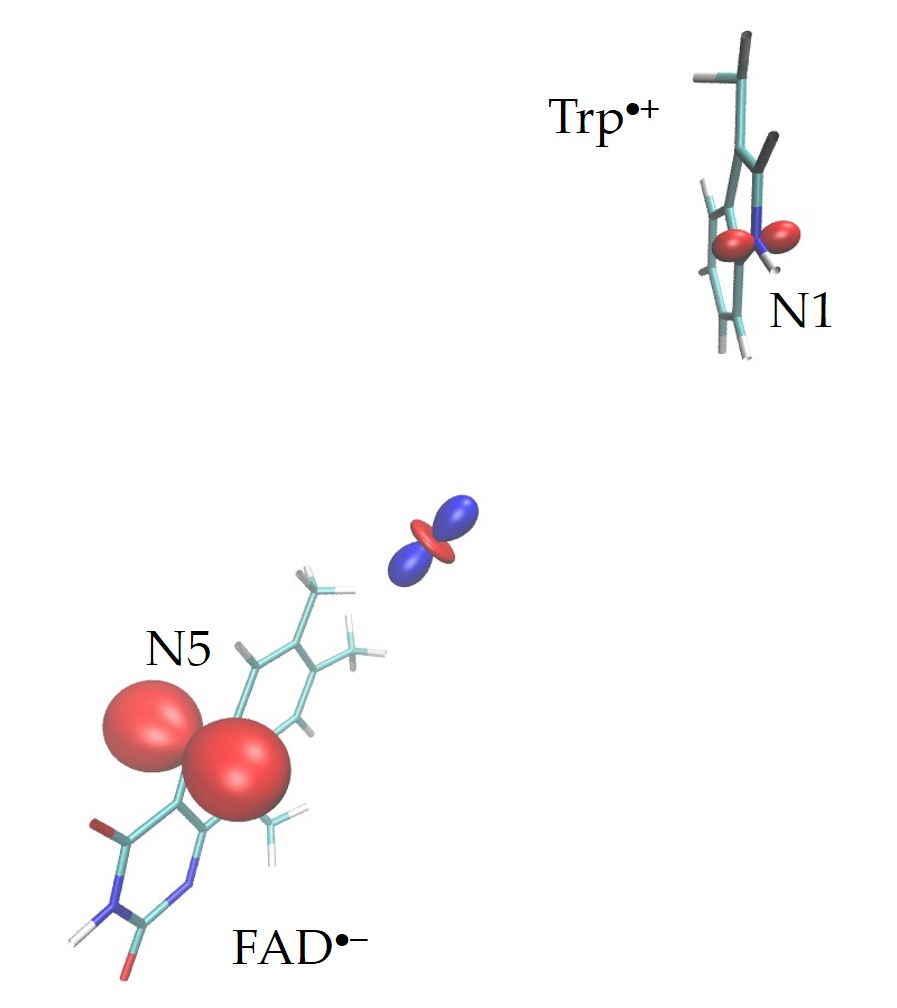}
    \caption*{(2)}
    \label{RP2}
\end{subfigure}
\hfill
\begin{subfigure}{0.24\textwidth}
    \includegraphics[width=\textwidth]{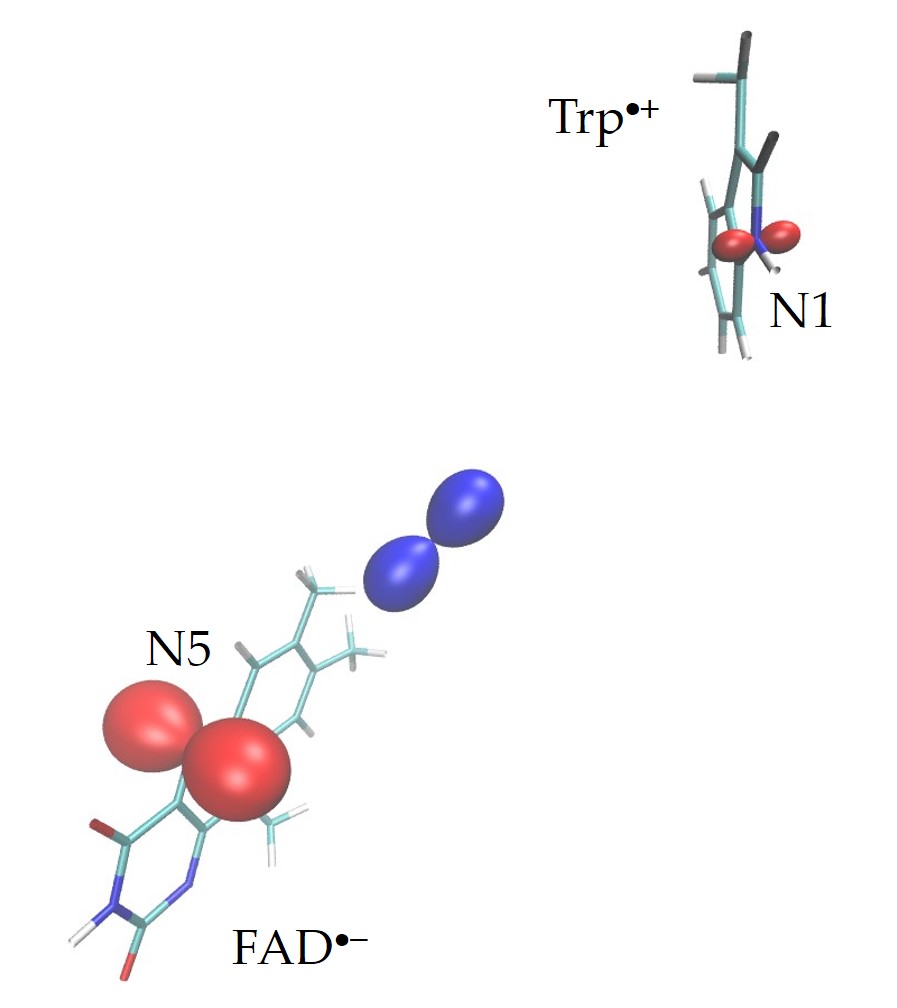}
    \caption*{(3)}
    \label{RP3}
\end{subfigure}
\hfill
\begin{subfigure}{0.24\textwidth}
    \includegraphics[width=\textwidth]{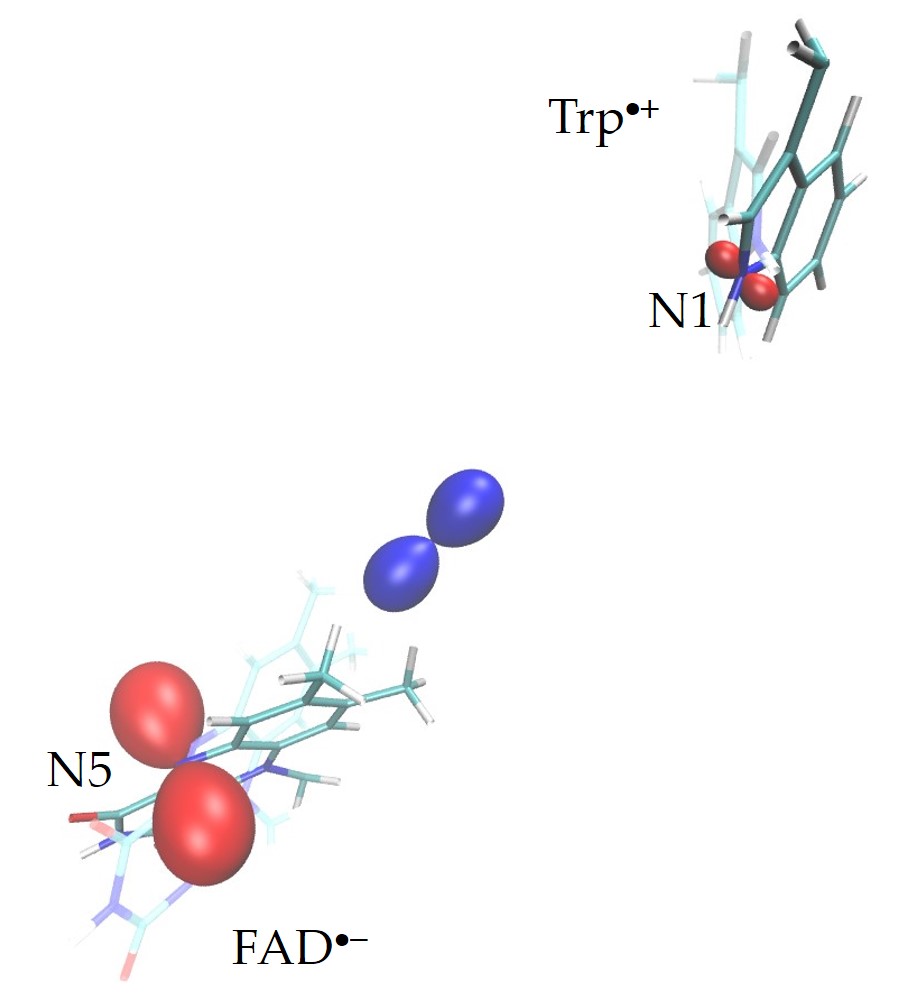}
    \caption*{(4)}
    \label{RP4}
\end{subfigure}
\caption{The arrangements of the FAD$^{\bullet{-}}$ and Trp$^{\bullet{+}}$ radicals and 
the spin couplings in the considered models of the cryptochrome radical pair. 
The surfaces display the shapes of the couplings (see the text). 
(1) RP1: the radical pair with the single HFC at each radical, N5 for FAD$^{\bullet{-}}$ 
and N1 for Trp$^{\bullet{+}}$. 
(2) RP2: the same structure and HFC interaction as in RP1, but with inclusion of the 
EED spin coupling. 
(3) RP3: the same structure as in RP1 and RP2, but with the electron exchange spin 
coupling added to the couplings of RP2. 
(4) RP4: the arrangement of the radicals and couplings derived from these arrangements 
and couplings in RP3 such that the HFC axis of each radical becomes orthogonal to 
the EED axis.}
\label{fig_1}
\end{figure}
}

For the model radical pairs RP1, RP2 and RP3, the mutual arrangement of the 
FAD$^{\bullet{-}}$ and Trp$^{\bullet{+}}$ coincides with that computed from the 
cryptochrome structure~\cite{Levy-2013-03} when deriving the HFC and EED tensors. 
For RP4, the radicals are oriented so that their HFC axes become orthogonal to the 
EED axis. Each radical is rotated around the average location of its unpaired electron, 
so that the EED tensor and hence the radical-pair EED axis remain unchanged.
%
% Table
%
\bigskip
{\linespread{1.2}
\begin{table}[h]%\footnotesize
\centering
\begin{tabular}{clll}
\hline
\multicolumn{4}{c}{Hyperfine coupling} \\
\hline
\multirow{6}{*}{RP1,RP2,RP3}
 &\multirow{3}{*}{FAD N5} 
  & $A_1=     -0.0799$ & ${\bf e}_1=(     -0.0843,\;\;\;0.0579,\;\;\;0.9948)$ \\
 && $A_2=     -0.0751$ & ${\bf e}_2=(     -0.0820,\;\;\;0.5635,     -0.1023)$ \\
 && $A_3=\;\;\;1.8338$ & ${\bf e}_3=(\;\;\;0.5665,\;\;\;0.8241,\;\;\;0\;\;\;\;\;\;\;\;)$ \\
\cline{2-4}
 &\multirow{3}{*}{Trp N1}
  & $A_1=     -0.0450$ & ${\bf e}_1=(\;\;\;0.7468,     -0.0457,\;\;\;0.6635)$ \\
 && $A_2=     -0.0345$ & ${\bf e}_2=(     -0.6623,\;\;\;0.0401,\;\;\;0.7481)$ \\
 && $A_3=\;\;\;0.5486$ & ${\bf e}_3=(     -0.0610,     -0.9981,\;\;\;0\;\;\;\;\;\;\;\;)$ \\ 
\hline
\multirow{6}{*}{RP4}
 &\multirow{3}{*}{FAD N5} 
  & $A_1=     -0.0799$ & ${\bf e}_1=(\;\;\;0.2314,\;\;\;0.0579,\;\;\;0.9069)$ \\
 && $A_2=     -0.0751$ & ${\bf e}_2=(\;\;\;0.5460,\;\;\;0.5635,\;\;\;0.1785)$ \\
 && $A_3=\;\;\;1.8338$ & ${\bf e}_3=(     -0.8052,\;\;\;0.8241,\;\;\;0.3817)$ \\ 
 \cline{2-4}
 &\multirow{3}{*}{Trp N1}
  & $A_1=     -0.0450$ & ${\bf e}_1=(     -0.8111,     -0.2132,\;\;\;0.5447)$ \\
 && $A_2=     -0.0345$ & ${\bf e}_2=(     -0.6623,\;\;\;0.7755,\;\;\;0.4106)$ \\
 && $A_3=\;\;\;0.5486$ & ${\bf e}_3=(     -0.0610,     -0.5942,\;\;\;0.7313)$ \\
\hline\hline
\multicolumn{4}{c}{EED spin coupling} \\
\hline
RP1 && $D_1=D_2=D_3=0$ & \\
\hline
\multirow{3}{*}{RP2,RP3,RP4} 
&& $D_1=     -0.8546$ & ${\bf e}_1=(     -0.1829,\;\;\;0.8023,\;\;\;0.5682)$ \\
&& $D_2=\;\;\;0.4273$ & ${\bf e}_2=(\;\;\;0.9559,\;\;\;0.2803,     -0.0881)$ \\
&& $D_3=\;\;\;0.4273$ & ${\bf e}_3=(\;\;\;0.2230,     -0.5270,\;\;\;0.8182)$ \\
\hline\hline
\multicolumn{4}{c}{Electron exchange spin coupling} \\
\hline
RP1, RP2 && $J=0$ & \\
\hline
RP3, RP4 && $J=-0.4273$ & \\
\hline
\end{tabular}
\caption{The spin couplings for RP1, RP2, RP3 and RP4. The principal values of the 
HFC and EED tensors and the values of the electron spin exchange constant are in mT. 
The principal directions of the tensors refer to the coordinate frame with the $z$-axis 
orthogonal to the HFC axes of RP1.}
\label{table}
\end{table}
}

Figure~\ref{fig_1} shows the arrangements of the radicals and the shapes of the spin 
couplings in the models, and Table~\ref{table} gives the values, principal values, and 
principal directions of the couplings. The shapes of the couplings are shown by the 
surfaces defined by the relation $d({\bf n}) = 3 T_{ik} n_i n_k$ [bohrs/mT]. 
Each surface corresponds to the distance $|d|$ from a center in the direction of a 
unit vector ${\bf n}$; the red color corresponds to positive and the blue color to 
negative values of d; $T_{ik} = A_{ik}$ for the hyperfine coupling; $T_{ik} = D_{ik}$ 
for the EED coupling; and $T_{ik} = D_{ik} + J\delta_{ik}$ for the EED and electron 
exchange coupling, with $\delta_{ik}$ being the Kronecker delta. We set the surface 
centers at the respective nucleus sites for the hyperfine couplings, and at the midpoint 
between the average locations of the unpaired electrons of the radicals for the EED and 
electron exchange coupling. The plots in Figure~\ref{fig_1} and the tensor principal 
directions in Table~\ref{table} refer to the coordinate system with the $z$-axis 
orthogonal to the HFC axes of both radicals in the RP1, RP2 and RP3 models; this system 
is displayed in Figure~\ref{fig_2}(a).

For each model, we start analyzing the field-direction response by computing the 
eigenstates of the spin Hamiltonian~(\ref{H}). We employ a standard numerical 
diagonalization of a Hamiltonian matrix represented in the basis set of the spin states 
of individual spins. The eigenstates $|n\rangle$ (denoted by the spin-level number $n$) 
provide a complete quantum-mechanical description of an isolated, non-decaying, 
radical pair. For each state, we determine the singlet weight 
$S_n = \left\langle n \left| P^{\rm S} \right| n \right\rangle$, where $P^{\rm S}$ 
is an operator projecting on the singlet state of the radical pair. The singlet weights 
show that, generally, due the hyperfine coupling, neither state is a pure singlet 
($S_n = 1$) or triplet ($S_n = 0$) state. In addition, the weights depend on the field 
direction, which reflects the role of the separate states in the compass effect.

Resulting from the variety of singlet weights, the electron spin character of the 
radical pair changes with time. The singlet and triplet parts of the isolated radical 
pair oscillate around time-independent values $\langle\Phi_{\rm S}\rangle$ and 
$\langle\Phi_{\rm T}\rangle$ with the sum 
$\langle\Phi_{\rm S}\rangle + \langle\Phi_{\rm T}\rangle = 1$. We refer to 
$\langle\Phi_{\rm S}\rangle$ and $\langle\Phi_{\rm T}\rangle$ as the interconversion 
singlet and triplet yields, respectively, and consider their dependence on the field 
direction as an output of the compass mechanism. For a singlet-born radical pair, 
it is practical to calculate 
\begin{equation}
\langle\Phi_{\rm S}\rangle 
= Z^{-1} \sum_{\substack{n,n'\\(E_n=E_{n'})}}
         \left\langle n  \left| P^{\rm S} \right| n' \right\rangle
         \left\langle n' \left| P^{\rm S} \right| n  \right\rangle,
\label{isy_1}
\end{equation}
where $Z$ is the nuclear spin multiplicity, and then determine 
$\langle\Phi_{\rm T}\rangle = 1 - \langle\Phi_{\rm S}\rangle$. The sum in the 
formula~(\ref{isy_1}) includes only the terms for which the energies $E_n$ and $E_{n'}$ 
are equal. When all the states are non-degenerate, such terms are those with $n = n'$, 
leading to
\begin{equation}
\langle\Phi_{\rm S}\rangle = Z^{-1} \sum_n S_n^2.
\label{isy_2}
\end{equation}
The expressions~(\ref{isy_1}) and (\ref{isy_2}) are readily obtained within 
a standard quantum-mechanical approach~\cite{Bezchastnov-2023-01-20} to the 
spin properties solely determined by the eigenstates of the spin 
Hamiltonian~(\ref{H}). Equally, they can be derived using the spin density 
operator $\rho(t)$ (see, e.g., Refs.~\cite{Fay-2019-12-13,Timmel-1998-09-01}), 
which for the isolated radical pair obeys the von Neumann equation 
${\rm d}\rho/{\rm d}t = -{\rm i}(H\rho-\rho{H})$.

The spin density formalism is adopted for description of a transient radical pair, 
whose spin properties additionally depend on spin-selective 
reactions~\cite{Haberkorn-1976-11-01}. For example, the time evolution of the spin density 
operator, relevant to the radical-pair mechanism of magnetoreception, is commonly 
described by adding decay terms to the von Neumann 
equation~\cite{Hore-2016-07-05,Babcock-2020-04-02}
\begin{equation}
\frac{{\rm d}\rho}{{\rm d}t} = 
-{\rm i}\left(H\rho-\rho{H}\right)
-\frac{k_{\rm AB}}{2} \left( P^{\rm S} \rho + \rho P^{\rm S}\right) - k_{\rm C}\rho.
\label{rho}
\end{equation}

The first decay term is the rate of spin-selective recombination of the radical pair, 
which re-establishes the singlet state of the initial reactants A and B (unexcited 
FAD and Trp), whereas the second term is the rate of the spin-independent formation 
of the signaling state C (neutral protonated FAD radical), $k_{\rm AB}$ and $k_{\rm C}$ 
being the respective rate constants. Solving Eq.~(\ref{rho}) in an operator form of 
$\rho(t)$ is straightforward~\cite{Babcock-2020-04-02}. 
In practical calculations, we obtain the solution in the form of the matrix elements 
$\langle n |\rho(t)| n' \rangle$ for the singlet-born radical pair, 
$\langle n |\rho(0)| n' \rangle = Z^{-1} \langle n | P^{\rm S} | n' \rangle$, 
and compute the yield 
$\Phi_{\rm C} = k_{\rm C} \int_0^\infty {\rm d}t \sum_n \langle n |\rho(t)| n \rangle$ 
of the signaling state. The direction of the magnetic field primarily affects the 
singlet-triplet interconversion and thus the spin-dependent recombination, 
so that it ultimately affects the signaling state formation, 
which is independent of spin but attenuated by the recombination. 
Thus, the influence of the field direction on the signaling state becomes more 
pronounced if the recombination rate exceeds the rate of formation of the 
signaling state. We perform the simulations for the values 
$k_{\rm AB}^{-1} = 8.1$~ms and $k_{\rm C}^{-1} = 57$~ms, significantly exceeding 
the microsecond time scale of the electron spin resonance at the $50~\mu$T magnetic 
field. In this way, we simulate a slow decay of the radical pair, for which the 
quantum-mechanical spin coherences last sufficiently long \cite{Hiscock-2016-04-26} 
for developing spikes in the field-direction dependence of the signaling yield.

\section{Results}
\label{Results}

\subsection{RP1 model}
\label{RP1_model}

The model radical pair RP1 has only the intra-radical spin interaction from the FAD N5 
and Trp N1 HFC contributions, see Figure~\ref{fig_1}(1), and is similar to the system 
elucidating the role of the axial HFC contributions~\cite{Hiscock-2016-04-26} 
in the quantum needle effect. Previously~\cite{Bezchastnov-2023-01-20}, we 
considered RP1 in order to gain a quantum mechanical insight into the anisotropy of 
the magnetic response of the individual radicals. In the present work, we consider RP1 
because its symmetry allows accurate 3D visualization of the spiky anisotropy of the 
compass output, which we further study with respect to extending and tuning the spin 
interaction. We show 3D magnetic anisotropy in terms of the interconversion triplet 
yield, for which the spikes, for the singlet-born radical pair, develop as sharp 
enhancements captured in a polar graph better than sharp attenuations of the previously 
considered singlet yield. As a measure of the anisotropy, we calculate the parameter 
$\Gamma  = \left(\Phi_{\rm max}-\Phi_{\rm min}\right)/
           \left(\Phi_{\rm max}+\Phi_{\rm min}\right)$, 
expressed in percent, where $\Phi_{\rm max}$ and $\Phi_{\rm min}$ are the maximal and 
minimal values, respectively, of the interconversion triplet yield determined over the 
whole range of the relative orientation of the radical pair and magnetic field.
%
% Figure 2
%
\bigskip
{\linespread{1.2}
\begin{figure}[h]
\centering
\begin{subfigure}{0.49\textwidth}
    \includegraphics[width=\textwidth]{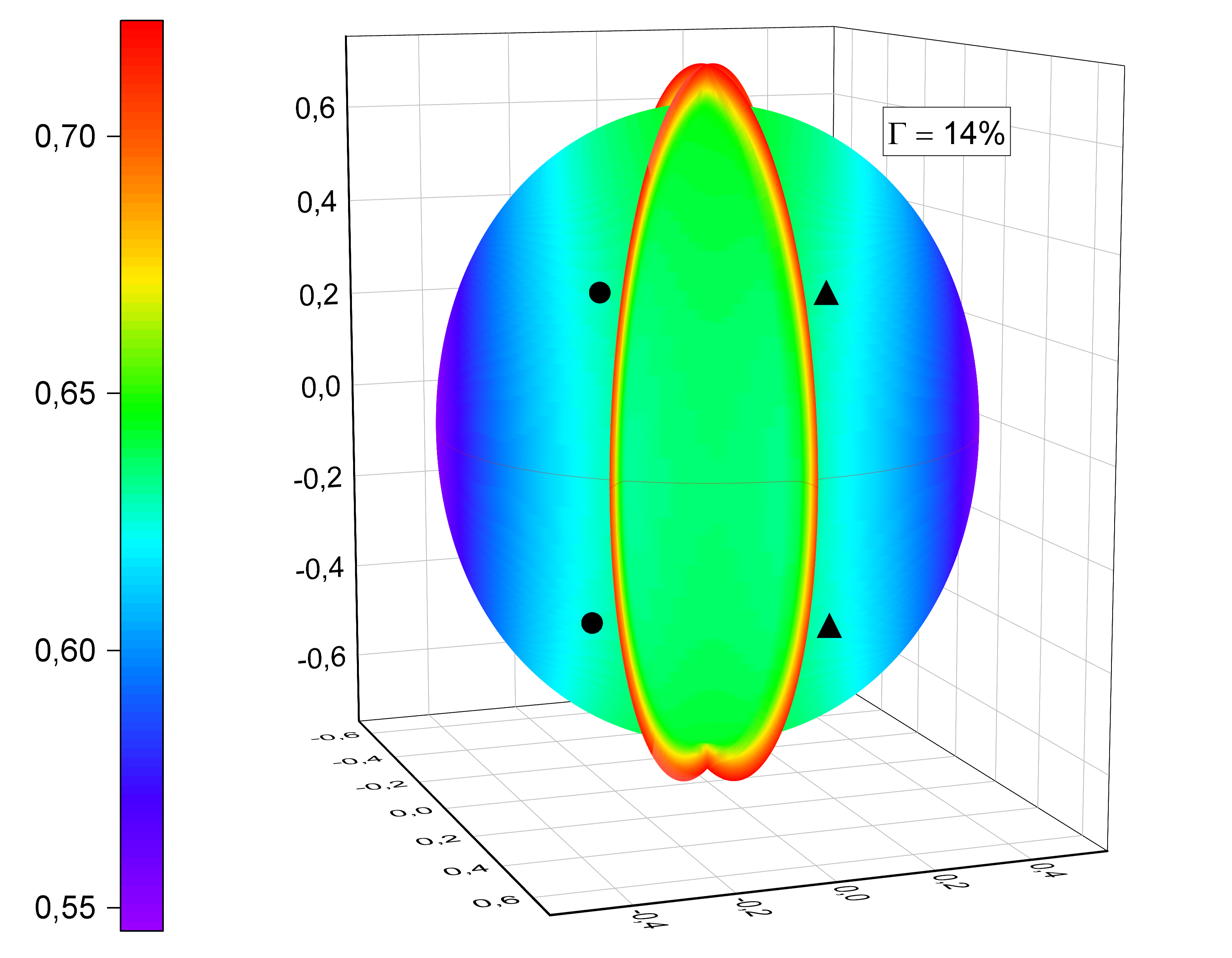}
    \caption{}
    \label{RP1a}
\end{subfigure}
\hfill
\begin{subfigure}{0.49\textwidth}
    \includegraphics[width=\textwidth]{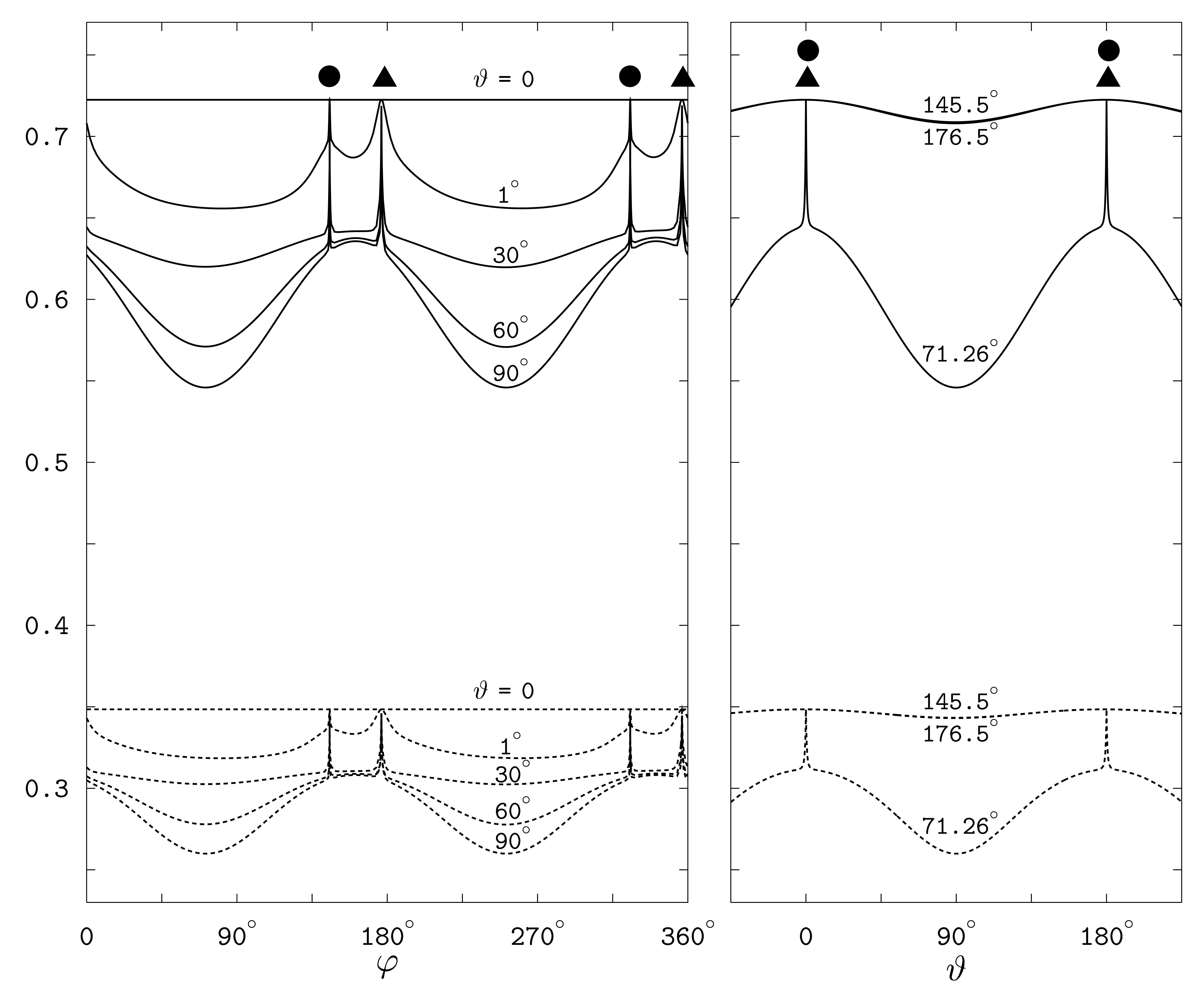}
    \caption{}
    \label{RP1b}
\end{subfigure}
\caption{The response of RP1 to the direction of the geomagnetic field. The symbols 
$\CIRCLE$ and $\blacktriangle$ indicate the spikes at the field directions orthogonal 
to the HFC axes FAD N5 and Trp N1, respectively. 
(a) The direction dependence of the interconversion triplet yield as a 3D polar graph. 
The direction of the magnetic field is defined in the coordinate frame shown in the plot. 
For the magnetic field pointing from the origin to the 3D surface, the distance from 
the origin to the surface gives the yield value according to the color scale. The 
vertical axis of the coordinate frame (the $z$-axis) is orthogonal to the HFC axes of 
the radicals, cf. Figure 1(1). 
(b) The dependence of the triplet yield (solid lines) and the yield of the signaling 
state (dashed lines) on the magnetic field direction as 2D graphs. 
Left plot: the yields varying with the azimuthal angle $\varphi$ for the selected 
values (indicated near the curves) of the polar angle $\vartheta$, with $\varphi$ 
and $\vartheta$ defining the magnetic field direction in the coordinate frame of the 
polar graph in Figure 2(a). 
Right plot: the yields varying with $\vartheta$ for the $\varphi$ values indicated 
near the curves.}
\label{fig_2}
\end{figure}
}

The 3D polar graph in Figure~\ref{fig_2}(a) shows the magnetic anisotropy of 14\%. 
The graph displays the spikes as two thin rings; 
these are adjacent to the surface that reflects a smooth variation of the 
interconversion triplet yield with the magnetic direction. Each of the rings 
corresponds to the plane perpendicular to the HFC axis of either FAD N5 or Trp N1. 
For accurate calculations and visualization of the spikes, we choose a coordinate 
frame with the $xy$-plane defined by the HFC axes. In this frame, the planes of the spikes 
are orthogonal to the $xy$-plane, and their intersection is parallel to the $z$-axis. 
For the azimuthal $\varphi$ and polar $\vartheta$ angles defining the magnetic direction 
in this coordinate system, the positions of the spikes are determined by four values 
of $\varphi$ independent of $\vartheta$. Explicitly, for any polar angle 
$0 \leq \vartheta \leq 180^\circ$, the spikes arise at the values of the azimuthal angle 
$\varphi_1 = 145.5^\circ$ and $\varphi_2 = 325.5^\circ$ for the FAD N5 spikes, and 
$\varphi_3 = 176.5^\circ$ and $\varphi_4 = 356.5^\circ$ for the Trp N1 spikes. 
The pairs $\varphi_1$, $\varphi_2$ and $\varphi_3$, $\varphi_4$ correspond to the 
opposite directions of the magnetic field in the planes of the spikes, so that 
$\varphi_2 - \varphi_1 = \varphi_4 - \varphi_3 = 180^\circ$. The difference 
$\varphi_3 - \varphi_1 = \varphi_4 - \varphi_2 = 31^\circ$ is the value of the angle 
between the FAD N5 and Trp N1 HFC axes. To produce the polar plots, we used the 
OriginPro (OriginLab Corporation, Northampton, MA, USA) software package, 
which required a rectangular grid over the polar and azimuthal angles. 
The above definition of the spike positions allowed a flexible 
adjustment of the grid density to the spikes: the $\vartheta$ values were distributed 
over the grid uniformly, while the $\varphi$ values were varied in appropriately small 
increments around the spike positions and in much larger increments between them. 
We also exploited the symmetry of the yield with respect to the inversion of the 
magnetic field direction by computing the triplet yield only for the upper hemisphere 
of the directions and assigning the computed values to the directions in the bottom 
hemisphere. Note that in the coordinate frame used for constructing the polar graph, 
the rings of spikes define the meridional planes crossing each other along the $z$-axis; 
in Figure~\ref{fig_2}(a) the rings are labeled $\CIRCLE$ and $\blacktriangle$ for the 
FAD N5 and Trp N1 related spikes, respectively. We regard the crossing of the spike rings 
as the {\em special direction} of the magnetic field 
noted previously~\cite{Bezchastnov-2023-01-20} and determined by the anisotropy of 
the hyperfine interaction. 

Figure~\ref{fig_2}(b) shows the interconversion triplet yield and the yield of the 
signaling state varying with one of the angles, $\varphi$ or $\vartheta$, at certain 
values of the other angle. The $\varphi$-profiles of the yields display spikes at four 
positions, $\varphi_1$, $\varphi_2$, $\varphi_3$ and $\varphi_4$, which are independent 
of $\vartheta$: see the left plot of the figure. The exception is the $\varphi$-profiles 
for $\vartheta = 0$ that collapse to the constant yield values corresponding to the 
magnetic direction towards the upper pole in the polar graph of Figure~\ref{fig_2}(a), 
where the two rings of spikes cross each other. Note that the spike maximum values are 
the same for all $\varphi$-profiles, and coincide with the values at the spike crossing 
at the pole $\vartheta = 0$. The yield minima between the spikes become deeper as the 
$\vartheta$ value approaches $90^\circ$, corresponding to the magnetic directions in the 
equatorial plane of the polar graph. 

The right plot of Figure~\ref{fig_2}(b) shows the 
$\vartheta$-profiles of the yields for three values of $\varphi$: 
$\varphi = 145.5^\circ$ and $\varphi = 156.5^\circ$ define the positions of the FAD N5 
and Trp N1 related spikes, respectively, and $\varphi = 71.26^\circ$ corresponds to the 
deepest minimum of the interconversion triplet yield. The respective profiles describe 
the magnetic anisotropy in the planes of the spike rings of the 3D polar graphs, 
and in the meridional plane between the rings. The $\vartheta$-profiles of two rings 
practically coincide with each other and reflect smooth variations of the spike 
intensities between maximal and minimal values approached as the magnetic field vector 
aligns with the directions of the poles ($\vartheta = 0$ and $\vartheta = 180^\circ$) 
and with the equatorial plane ($\vartheta = 90^\circ$) of the polar graph, respectively. 
For the azimuthal angle of the deepest minimum, the yields vary with the polar angle 
over the largest ranges of values and exhibit spikes at $\vartheta = 0$ and 
$\vartheta=180^\circ$, each of which results from merging the FAD N5 and Trp N1 spikes 
into one. The spikes are remarkably narrow, with an angular width of about $1^\circ$.
%
% Figure 3
%
\bigskip
{\linespread{1.2}
\begin{figure}[h]
\centering
\begin{subfigure}{0.32\textwidth}
    \includegraphics[width=\textwidth]{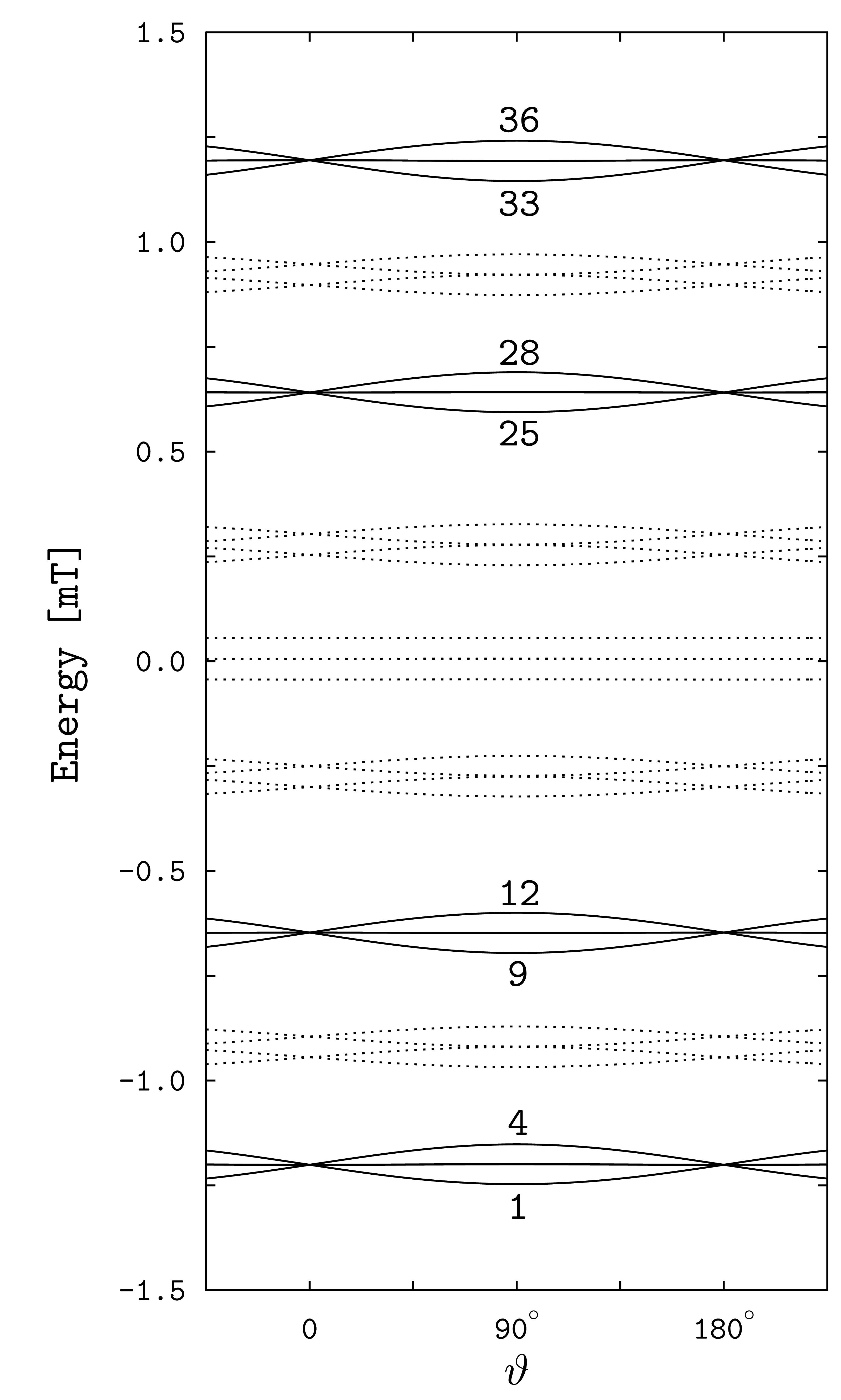}
    \caption{}
    \label{RP1_states_a}
\end{subfigure}
\hfill
\begin{subfigure}{0.32\textwidth}
    \includegraphics[width=\textwidth]{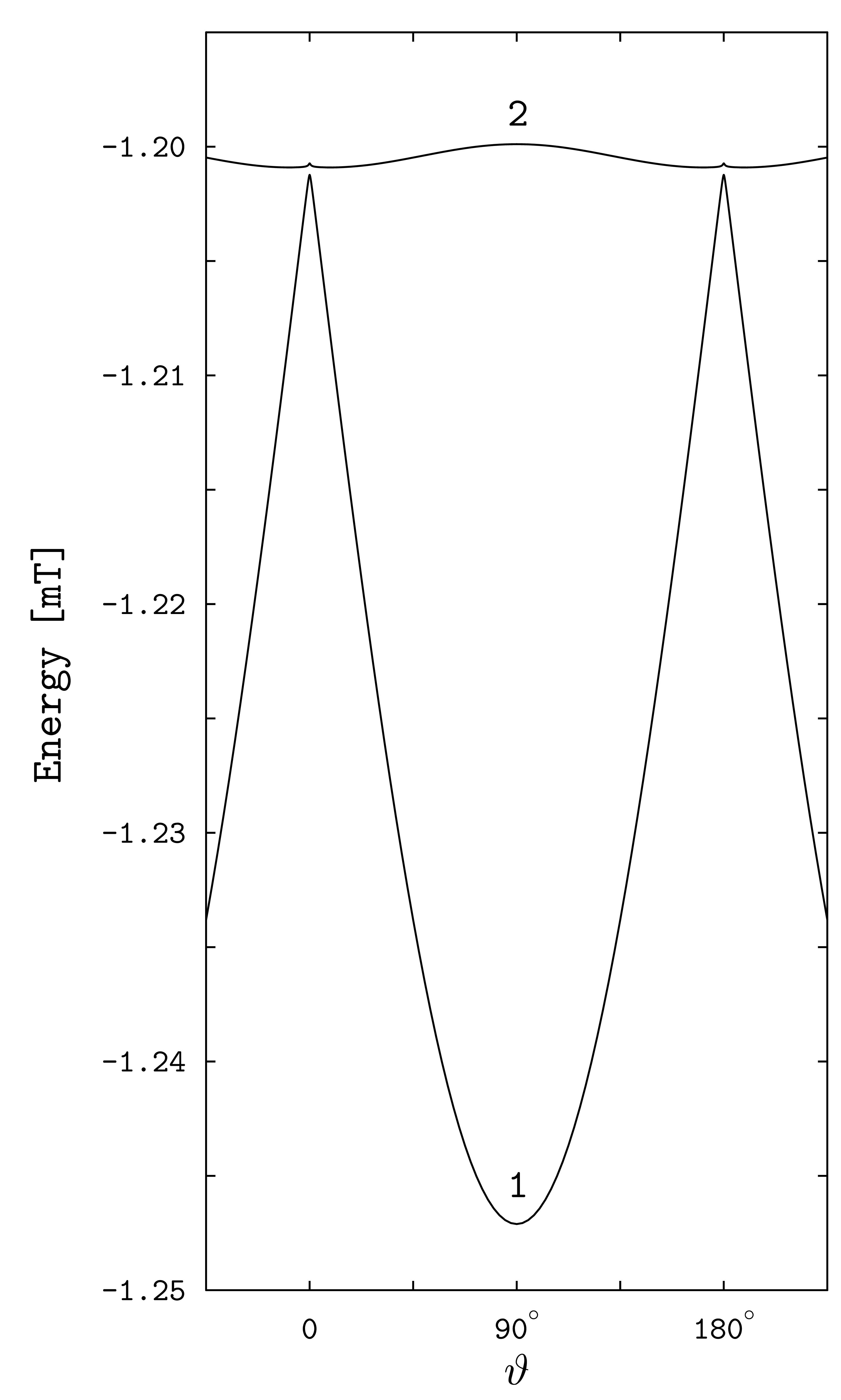}
    \caption{}
    \label{RP1_states_b}
\end{subfigure}
\hfill
\begin{subfigure}{0.32\textwidth}
    \includegraphics[width=\textwidth]{fig_3c.jpg}
    \caption{}
    \label{RP1_states_c}
\end{subfigure}
\caption{The energies and singlet weights of the spin states responsible for the sharp 
magnetic anisotropy of the RP1 model. 
(a) The dependence of the spin-state energies of the model radical pair on the polar 
angle $\vartheta$ at the azimuthal angle $\varphi=71.26^\circ$. The solid lines show 
the bunches of the energy levels responsible for the spikes in the yield profiles at 
$\vartheta=0$ and $\vartheta=180^\circ$, and the dashed lines show the other levels of 
in a total of 36 spin states. The numbers enumerate the levels in order of increasing 
energy. (b) The avoided crossing of the levels 1 and 2 at the positions of spikes. 
(c) The singlet weights of the levels 1 and 2. Note sharp variation of the weights at 
the spike positions.}
\label{fig_3}
\end{figure}

The spikes in the magneto-response anisotropy relate to properties of the spin states of 
the radical pair. We illustrate this in Figure~\ref{fig_3} in the variation of the 
quantum-state energies and singlet weights with the field direction. 
The energy levels of the states responsible for the spikes group in four bunches of four 
levels each, 1 to 4, 9 to 12, 25 to 28, and 33 to 36 (we enumerate the levels in the 
order of increasing energy): see Figure~\ref{fig_3}(a). The pairs of levels of each 
bunch, for example, the pairs (1,2) and (3,4), exhibit avoided crossings at the spike 
positions $\vartheta = 0$ and $\vartheta = 180^\circ$: see Figure~\ref{fig_3}(b) showing 
the level pair (1,2). The singlet weights in each pair are close to either the zero or 
$0.5$ value, except for the narrow angular domains of the avoided energy-level crossings, 
where the singlet weights sharply vary and approach each other: see Figure~\ref{fig_3}(c) 
demonstrating the singlet weights of the spin states 1 and 2. Previously, the 
study~\cite{Hiscock-2016-04-26} linked the spikes to the avoided crossings of 
the spin-energy levels, and the study~\cite{Bezchastnov-2023-01-20} related the 
sharp variations of the respective singlet weights to the flips in alignment of the 
unpaired electron spin to the HFC axis in each radical.

\subsection{RP2 model}
\label{RP2_model}

The model radical pair RP2 differs from RP1 only in the addition of the inter-radical 
EED spin coupling determined by the axial trace-less tensor ($D_1 = -2D_2 = -2D_3$, 
cf. Table~\ref{table}). However, the additional coupling, the form of which is shown 
in Figure~\ref{fig_1}(2), drastically modifies the response of the radical pair to the 
magnetic direction. Figure~\ref{fig_4} demonstrates the disappearance of the spikes, 
as well as the almost completely vanished anisotropy of the interconversion triplet 
yield and of the yield of the signaling state. The plots in the figure refer to the 
same coordinate system as in Figure~\ref{fig_2}. Note the narrow range of variation of 
the interconversion triplet yield corresponding to the anisotropy of 0.6\% of 
3D polar graph in Figure~\ref{fig_4}(a), as well as the pairs of profiles fairly close 
to each other in Figure~\ref{fig_4}(b). We calculated the latter profiles fixing the 
non-varying angles at the values from the pairs $\varphi = 174.0^\circ$, 
$\vartheta = 5^\circ$ and $\varphi = 256.2^\circ$, $\vartheta = 90^\circ$, which define 
the directions of the magnetic field at which the interconversion triplet yield is 
maximal and minimal, respectively.
%
% Figure 4
%
\bigskip
{\linespread{1.2}
\begin{figure}[h]
\centering
\begin{subfigure}{0.49\textwidth}
    \includegraphics[width=\textwidth]{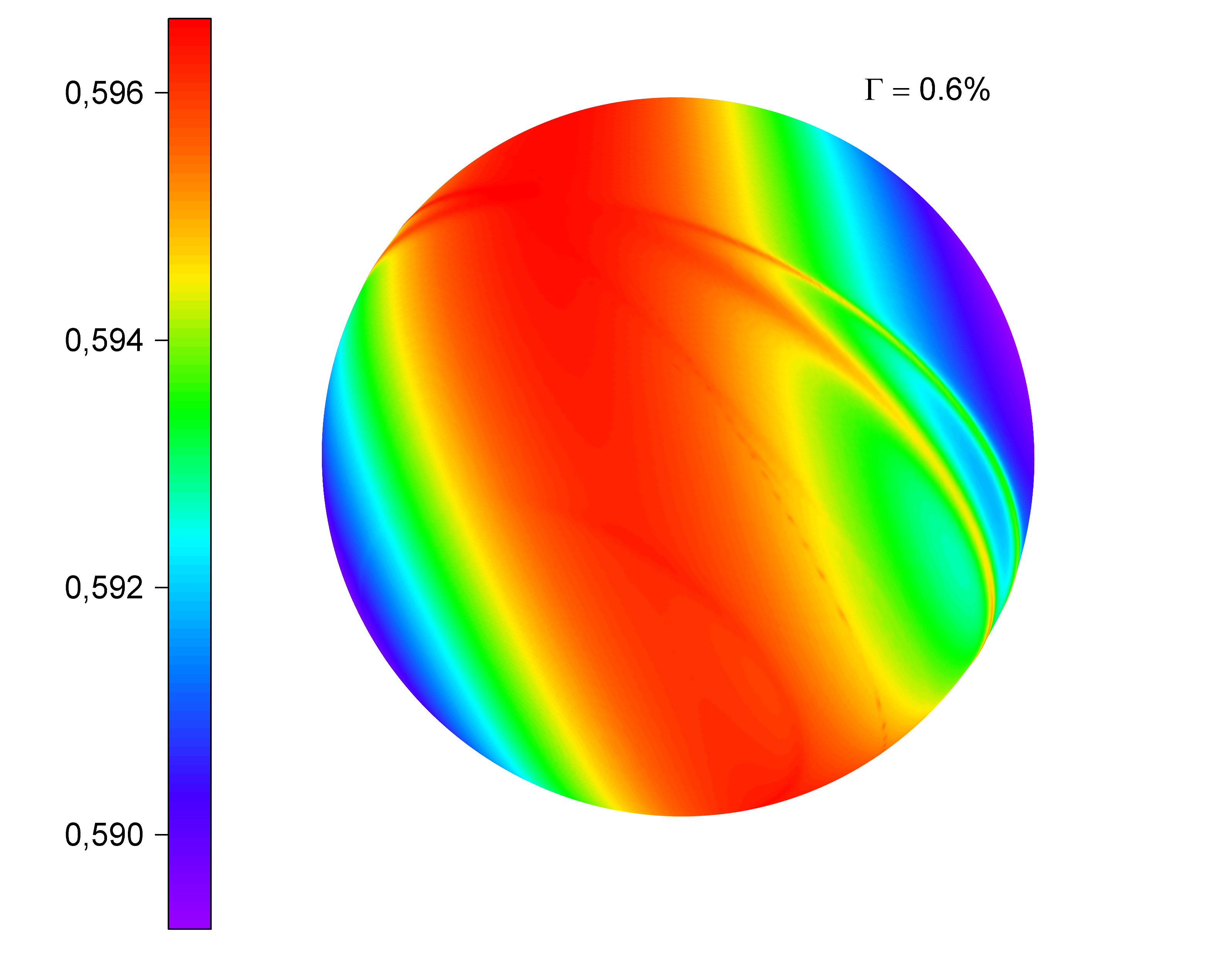}
    \caption{}
    \label{RP2a}
\end{subfigure}
\hfill
\begin{subfigure}{0.49\textwidth}
    \includegraphics[width=\textwidth]{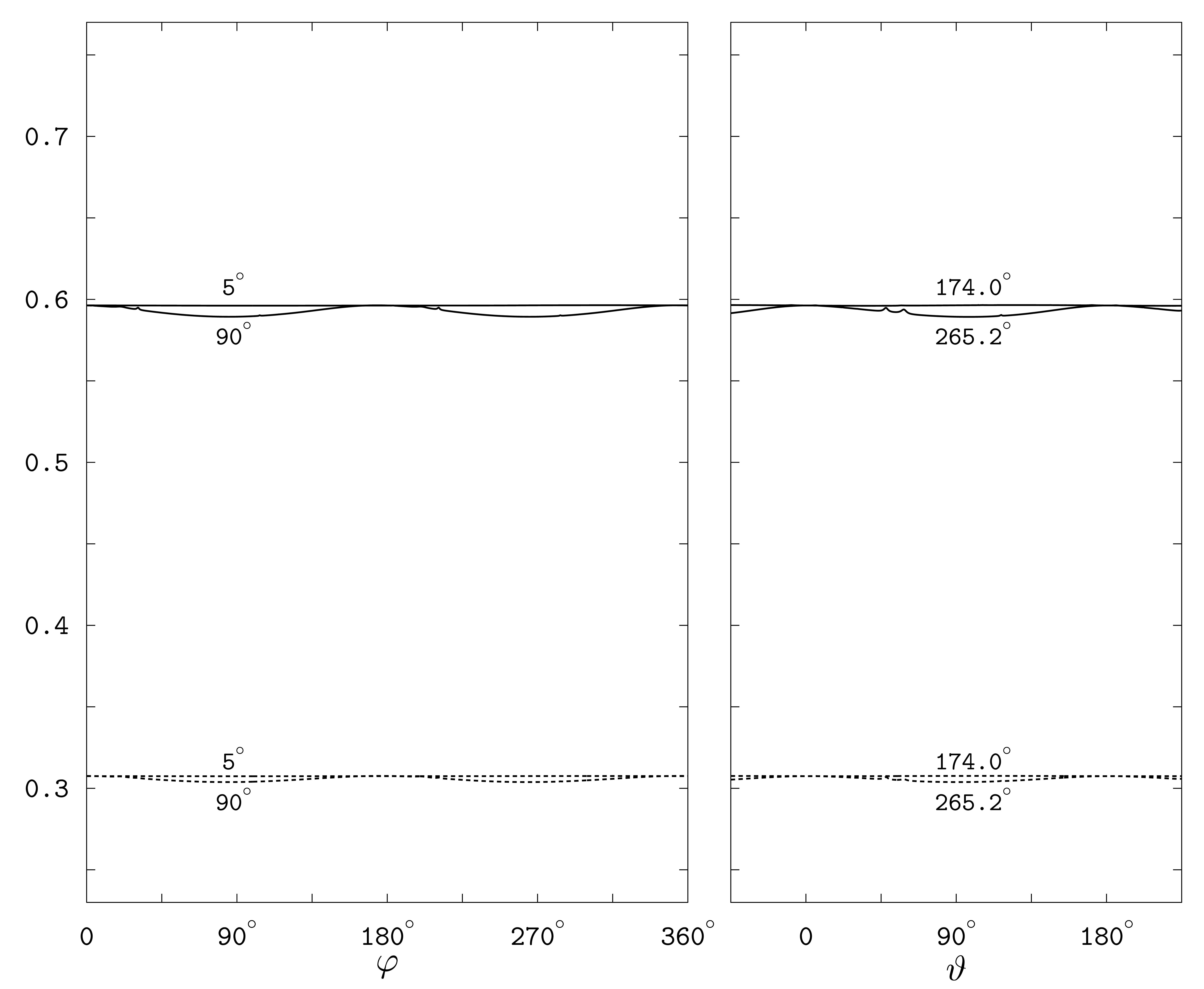}
    \caption{}
    \label{RP2b}
\end{subfigure}
\caption{The field-direction response of RP2. 
(a) The 3D polar graph for the interconversion triplet yield. 
(b) The triplet and signaling yields varying with one of the angles, $\vartheta$ or 
$\varphi$, at the selected values (indicated near the curves) of the other angle.}
\label{fig_4}
\end{figure}
}

In contrast to the spin states of the RP1 model, the RP2 states do not reflect the HFC 
anisotropy, and none of the singlet weights vary particularly sharply with the field 
direction. However, the RP2 model has states similar to the singlet and triplet states 
of a two-electron spin system subject to the EED spin interaction in the presence of 
a magnetic field. Note that the singlet (S) state of the latter two-electron system 
does not depend on the magnetic field. The triplet states, being decoupled from the 
singlet state, are affected by the field. In particular, the magnetic field couples 
the basis triplet states T$_{+1}$, T$_0$, and T$_{-1}$ defined according to the 
projection of the total electronic spin onto the EED axis. The coupling vanishes at 
the field alignment parallel to the EED axis, for which the triplet states 
transform into the basis states. Because of this property we refer to the triplet 
states in the presence of the magnetic field as the above-defined basis triplet states.
%
% Figure 5
%
\bigskip
{\linespread{1.2}
\begin{figure}[h]
\centering
\begin{subfigure}{0.32\textwidth}
    \includegraphics[width=\textwidth]{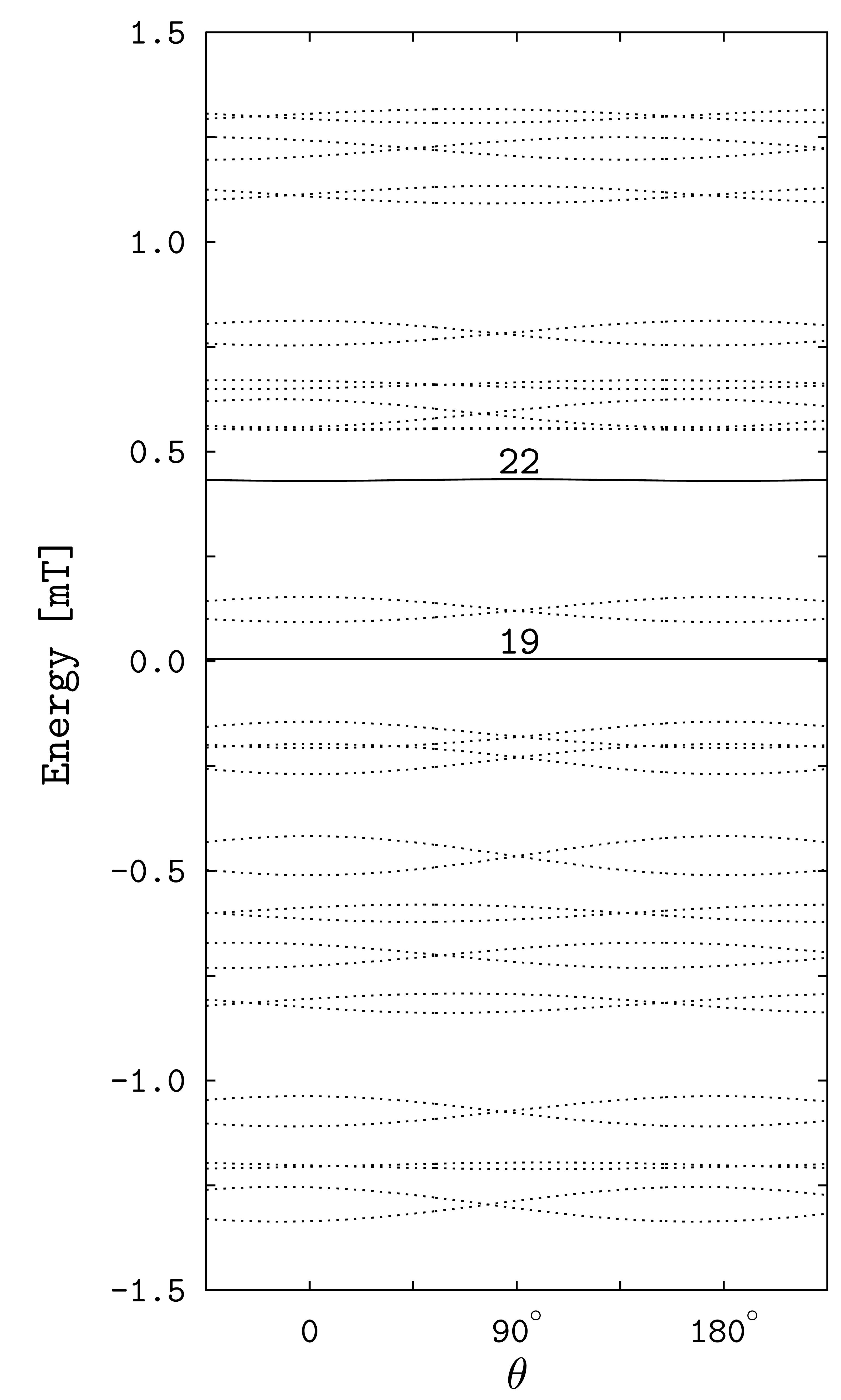}
    \caption{}
    \label{RP2_states_a}
\end{subfigure}
\hfill
\begin{subfigure}{0.32\textwidth}
    \includegraphics[width=\textwidth]{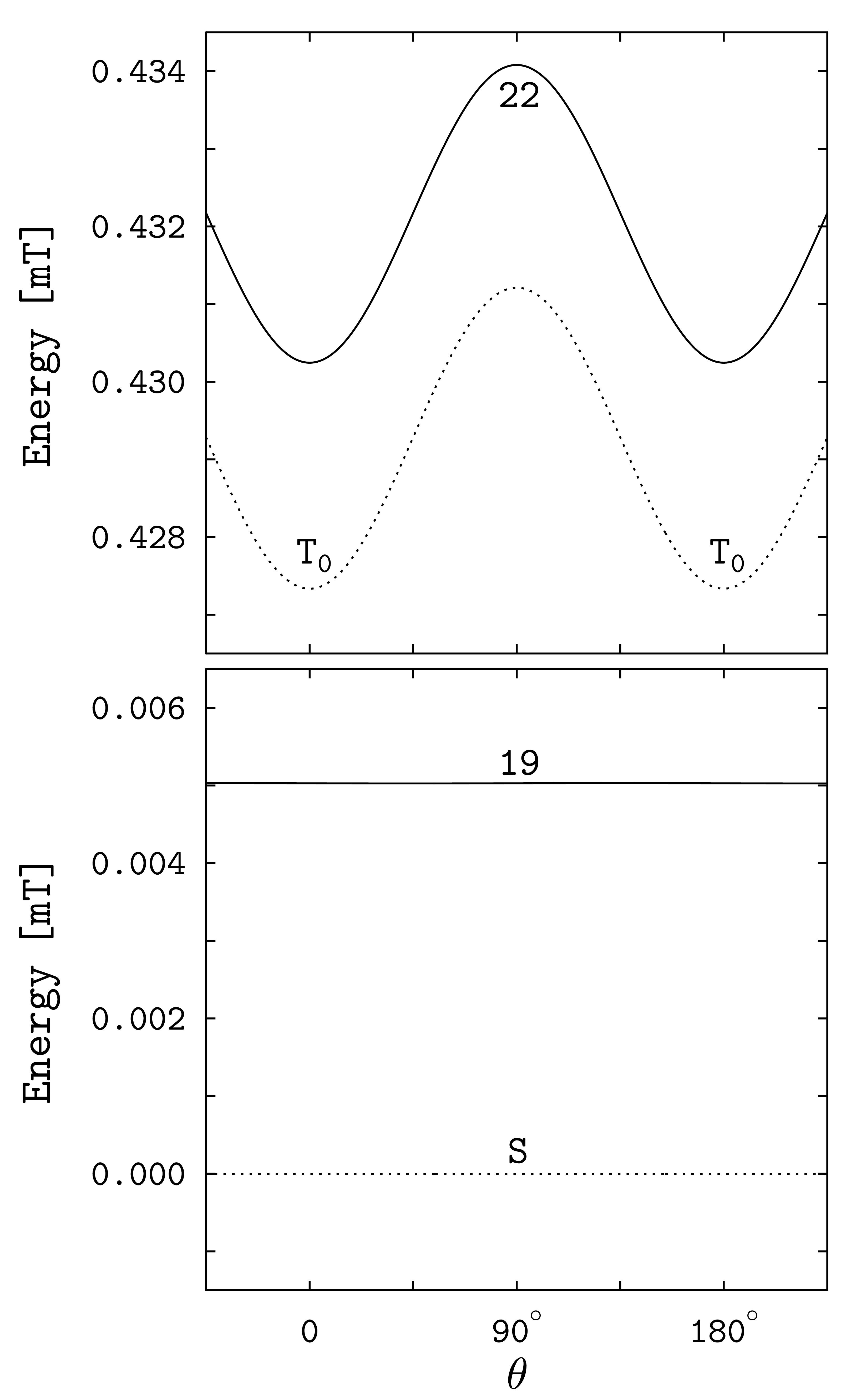}
    \caption{}
    \label{RP2_states_b}
\end{subfigure}
\hfill
\begin{subfigure}{0.32\textwidth}
    \includegraphics[width=\textwidth]{fig_5c.jpg}
    \caption{}
    \label{RP2_states_c}
\end{subfigure}
\caption{The spin-state properties of the RP2 model. 
(a) The spin-energy levels varying with the angle $\theta$ between the magnetic field 
and EED axis. The solid lines show the levels 19 and 22, and the dashed lines show the 
remaining 34 out of 36 levels. 
(b) The RP2 levels 19 and 22 (solid lines) compared to the singlet S and triplet 
T$_0$ levels of the electron spin pair (dashed lines). The notation T$_0$ reflects 
the definition of the triplet states with respect to the EED axis (see text). 
(c) The singlet weights of the spin states 19 and 22.}
\label{fig_5}
\end{figure}
}

For better comparison of the RP2 and two-electron spin states, we computed the state 
energies and singlet weights as functions of the angle $\theta$ between the magnetic 
field and the EED axis. Figure~\ref{fig_5} demonstrates that the RP2 states 19 and 22 
have energies very close to the energies of the two-electron states S and T$_0$, and 
the singlet weights of almost one and zero, respectively. The energy of the RP2 state 
19 does not depend on the field direction, and differs slightly from the singlet-state 
energy. For the RP2 state 22, the energy level follows the field-direction dependence 
pattern of the triplet T$_0$ level of the electron pair, with a somewhat smaller 
difference in values between the levels 22 and T$_0$ than between the levels 19 and S. 
The numerically calculated singlet $E_{\rm S}$ and triplet $E_{{\rm T}_0}$ energies 
agree with the values
\begin{align}
E_{\rm S} & = 0, && \mbox{independent of}~\theta,
\nonumber \\
E_{{\rm T}_0} & = 4\Delta, && \mbox{for}~\theta = 0~\mbox{or}~\theta = 180^\circ,
\label{energies_rp2}
\\
E_{{\rm T}_0} & = \Delta + \sqrt{(3\Delta)^2+B^2}, && \mbox{for}~\theta = 90^\circ,
\nonumber
\end{align}
where $\Delta = |D_1|/8$, obtained analytically for an axially symmetric EED spin 
coupling with a negative principal value corresponding to the symmetry axis 
($D_1 < 0$, $D_2 = D_3 = -D_1/2$, cf. Table~\ref{table}).

\subsection{RP3 model}
\label{RP3_model}

In the model radical pair RP3, we extended the inter-radical spin coupling to include the 
electron exchange contribution determined by the condition 
$-J = D_2 = D_3 = |D_1|/2$ (cf. Table~\ref{table}) that corresponds to a partial $J/D$ 
cancellation~\cite{Efimova-2008-03-01}. The tensor 
$T_{ik} = D_{ik} + J\delta_{ik}$, which couples the spins of unpaired electrons according 
to the last two terms of the Hamiltonian~(\ref{H}), became thereby an ``utmost axial'', 
with a single non-zero principal value corresponding to the direction of the EED axis: 
see Figure~\ref{fig_1}(3) for the tensor shape. When this tensor is the only source of 
the spin coupling, the S and T$_0$ states of two-electron spin system 
become degenerate~\cite{Efimova-2008-03-01}.
%
% Figure 6
%
\bigskip
{\linespread{1.2}
\begin{figure}[h]
\centering
\begin{subfigure}{0.49\textwidth}
    \includegraphics[width=\textwidth]{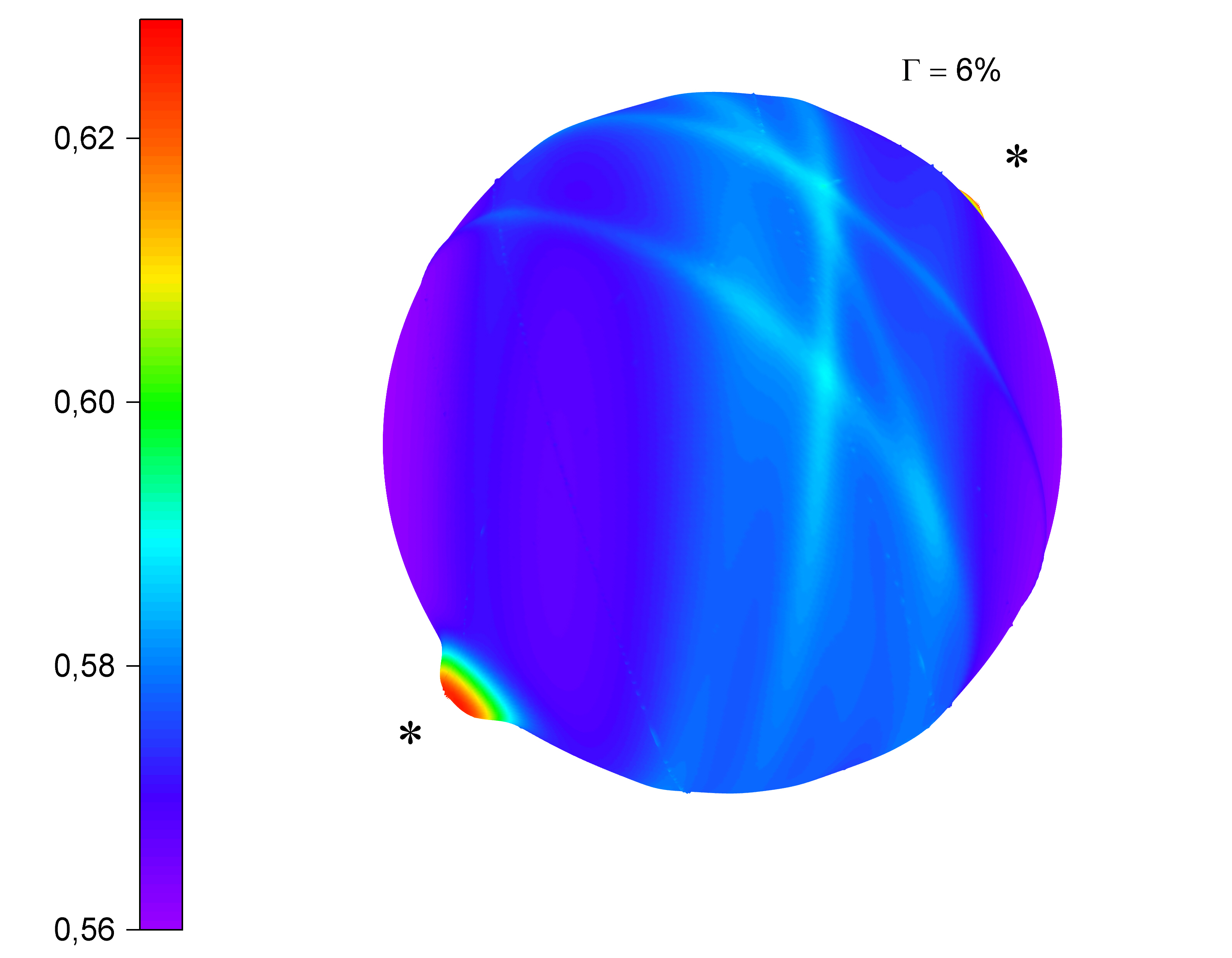}
    \caption{}
    \label{RP3a}
\end{subfigure}
\hfill
\begin{subfigure}{0.49\textwidth}
    \includegraphics[width=\textwidth]{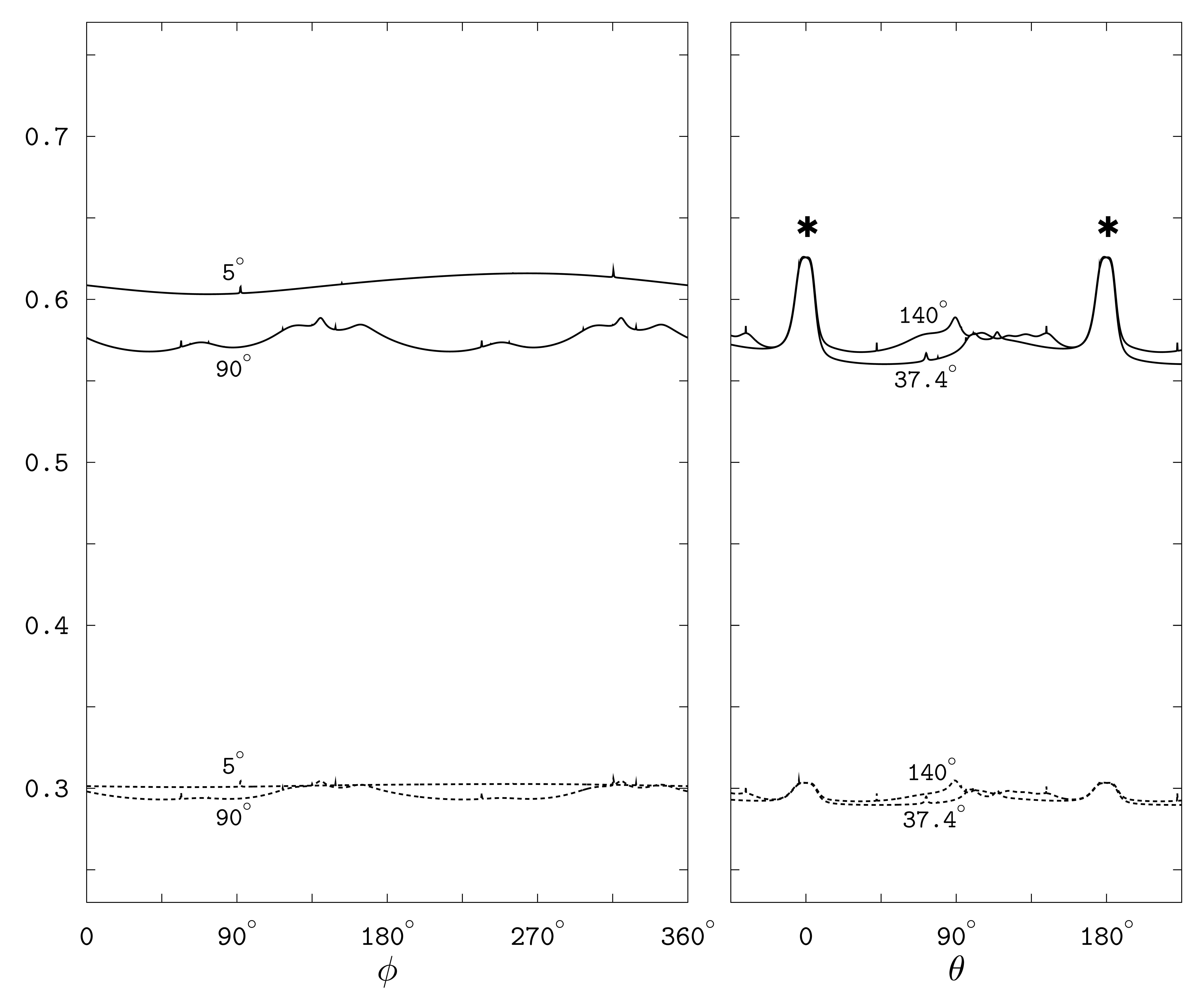}
    \caption{}
    \label{RP3b}
\end{subfigure}
\caption{The field-direction response of RP3. The symbols 
\raisebox{-0.45em}{\resizebox{!}{1.1em}{\bf *}} indicate the enhancements of the yields 
in the field directions parallel to the EED axis. 
(a) The magnetic anisotropy of the interconversion triplet yield as a 3D polar graph. 
The orientation of the surface refers to the coordinate system of Figure 2(a). 
(b) The dependence of the triplet and signaling yields on the azimuthal $\phi$, 
and polar, $\theta$, angles defined in a coordinate system with the $z$-axis aligned 
to the EED axis. The left plot shows the yields as functions of $\phi$ at the 
selected values of $\theta$ indicated near the curves. The right plot displays the 
dependencies of the yields on $\theta$ at constant $\phi$ (the values near the curves).}
\label{fig_6}
\end{figure}
}

The particular combination of the EED and exchange spin couplings in the RP3 model 
results in an anisotropy of the magnetic response emphasizing the direction of the 
EED axis. The interconversion triplet yield has the anisotropy of 6\% and 
is notably increased when the orientations 
of the magnetic field approach the directions parallel and anti-parallel to the EED axis: 
see the 3D polar graph in Figure~\ref{fig_6}(a) and the yield profiles along the angle 
$\theta$ between the magnetic field and EED axis in Figure~\ref{fig_6}(b). 
The enhancements develop on top of an almost isotropic background, within the angular 
range $\Delta\theta \approx 15^\circ$ around the yield maxima at $\theta = 0$ and 
$\theta = 180^\circ$. The yield of the signalling state displays a similar anisotropy, 
though with a less pronounced enhancement in the EED directions.

The S--T$_0$ degeneracy of the two-electron spin states is manifested in the spin states 
of the RP3 model. Figures~\ref{fig_7}(a) and \ref{fig_7}(b) demonstrate that the energy 
levels of the states 21 and 22 are very close to each other and similar to the S and 
T$_0$ levels of the two-electron spin system with the considered EED, exchange and 
Zeeman interactions. The lower-energy level 21 is almost independent of the magnetic 
direction, whereas the higher-energy level 22 is direction-modulated, closely 
approaching the lower level at the longitudinal orientations of the magnetic field 
relative to the EED axis, $\theta = 0$ and $\theta = 180^\circ$, and maximally 
diverging from the lower level at the transverse orientation, $\theta = 90^\circ$. 
For the lower level, only a slight direction dependence arises in narrow angular regions 
around the longitudinal field orientations, where the proximity of the two levels can be 
regarded as an {\em avoided crossing}. Except for the avoided crossings, the levels 21 
and 22 of the radical pair differ from the S and T$_0$ levels of the electron spin pair, 
respectively, by a small energy value. The energies $E_S$ and $E_{{\rm T}_0}$ can be 
calculated analytically,
\begin{align}
E_{\rm S} & = 3\Delta, && \mbox{independent of}~\theta,
\nonumber \\
E_{{\rm T}_0} & = 3\Delta, && \mbox{for}~\theta = 0~\mbox{or}~\theta = 180^\circ,
\label{energies_rp3}
\\
E_{{\rm T}_0} & = \sqrt{(3\Delta)^2+B^2}, && \mbox{for}~\theta = 90^\circ,
\nonumber
\end{align}
demonstrating that the S and T$_0$ levels are degenerate at the zero magnetic field and 
at the longitudinal orientations of a non-zero field, while the S level is not affected 
by the magnetic field.
%
% Figure 7
%
\bigskip
{\linespread{1.2}
\begin{figure}[h]
\centering
\begin{subfigure}{0.32\textwidth}
    \includegraphics[width=\textwidth]{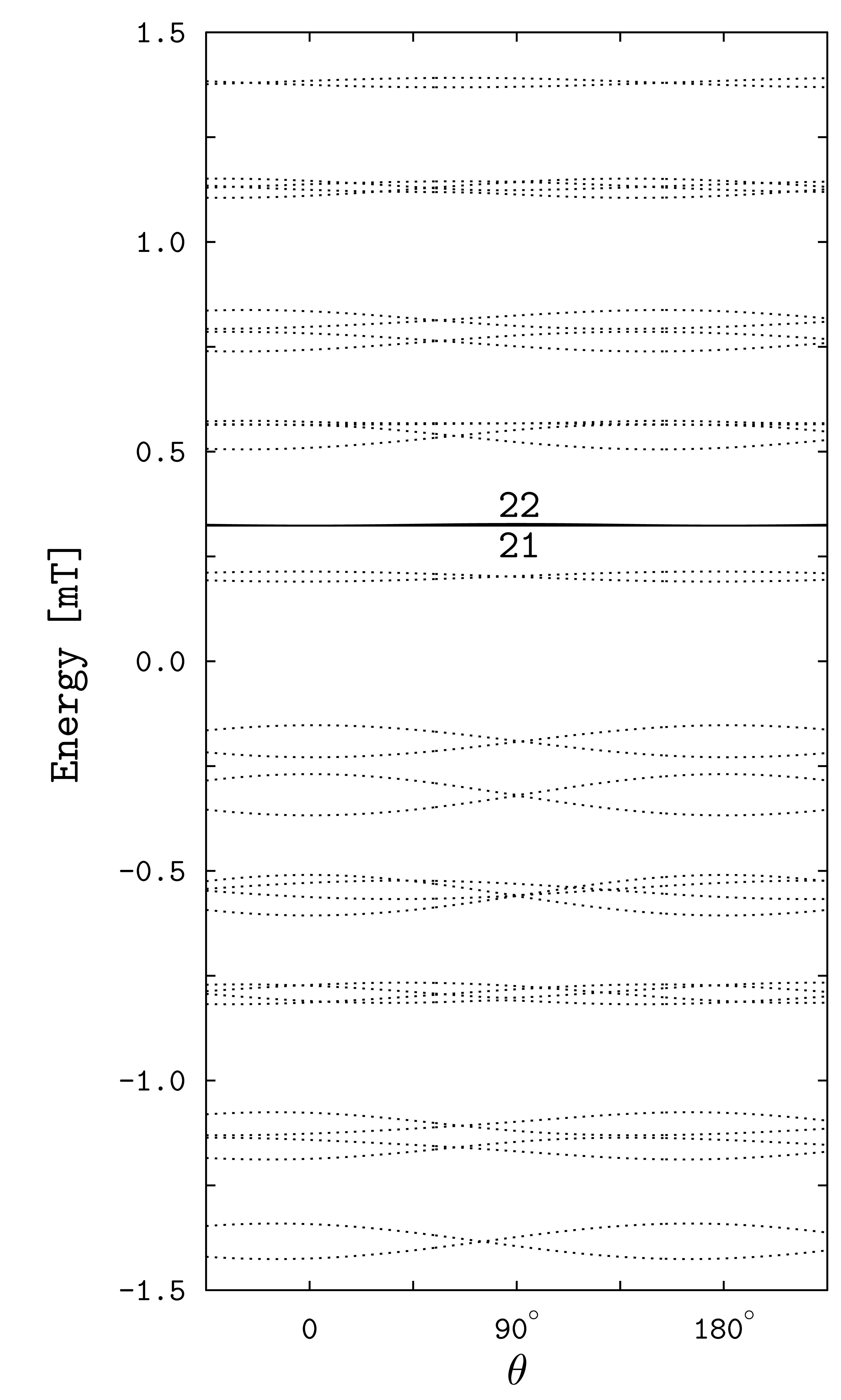}
    \caption{}
    \label{RP3_states_a}
\end{subfigure}
\hfill
\begin{subfigure}{0.32\textwidth}
    \includegraphics[width=\textwidth]{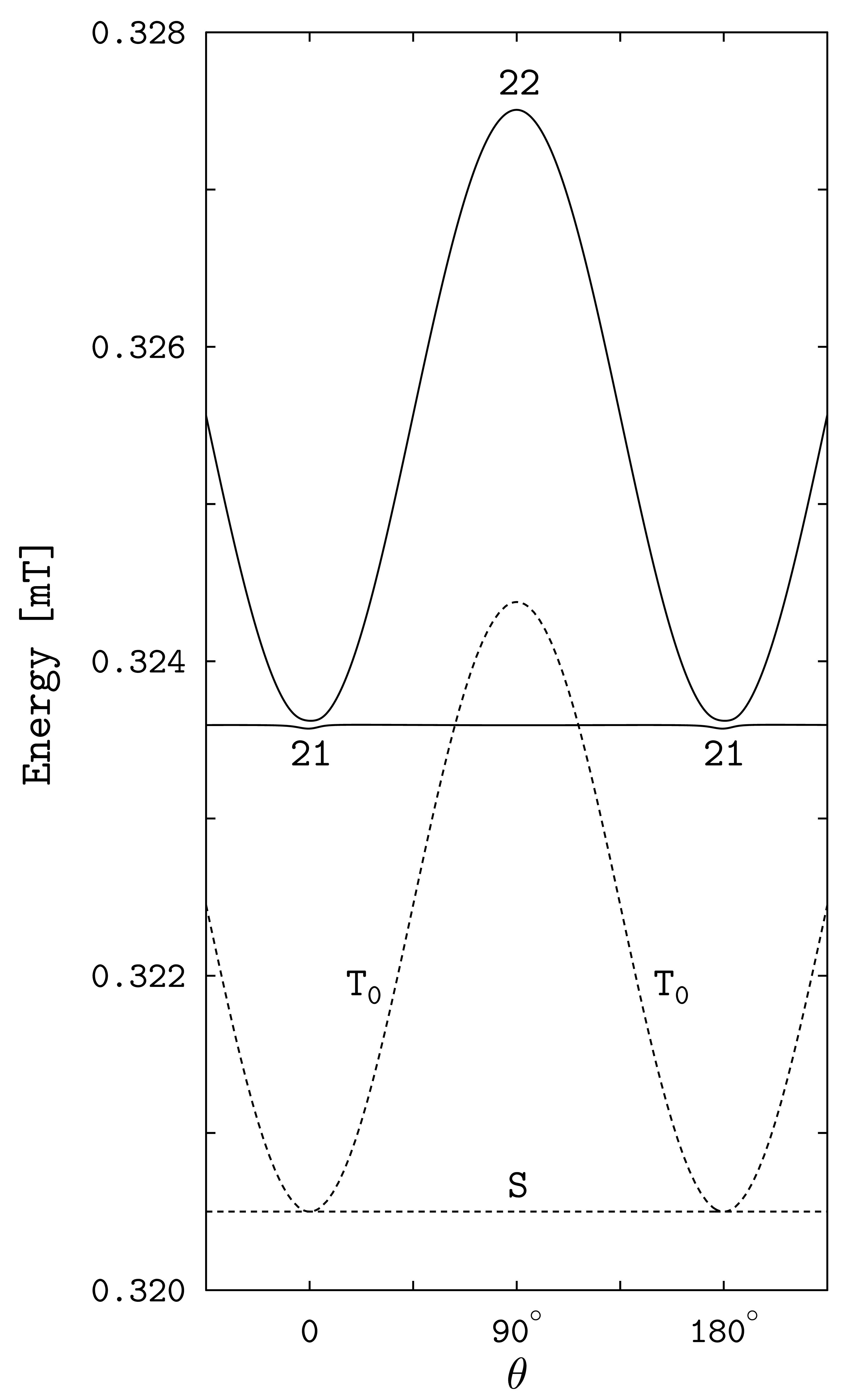}
    \caption{}
    \label{RP3_states_b}
\end{subfigure}
\hfill
\begin{subfigure}{0.32\textwidth}
    \includegraphics[width=\textwidth]{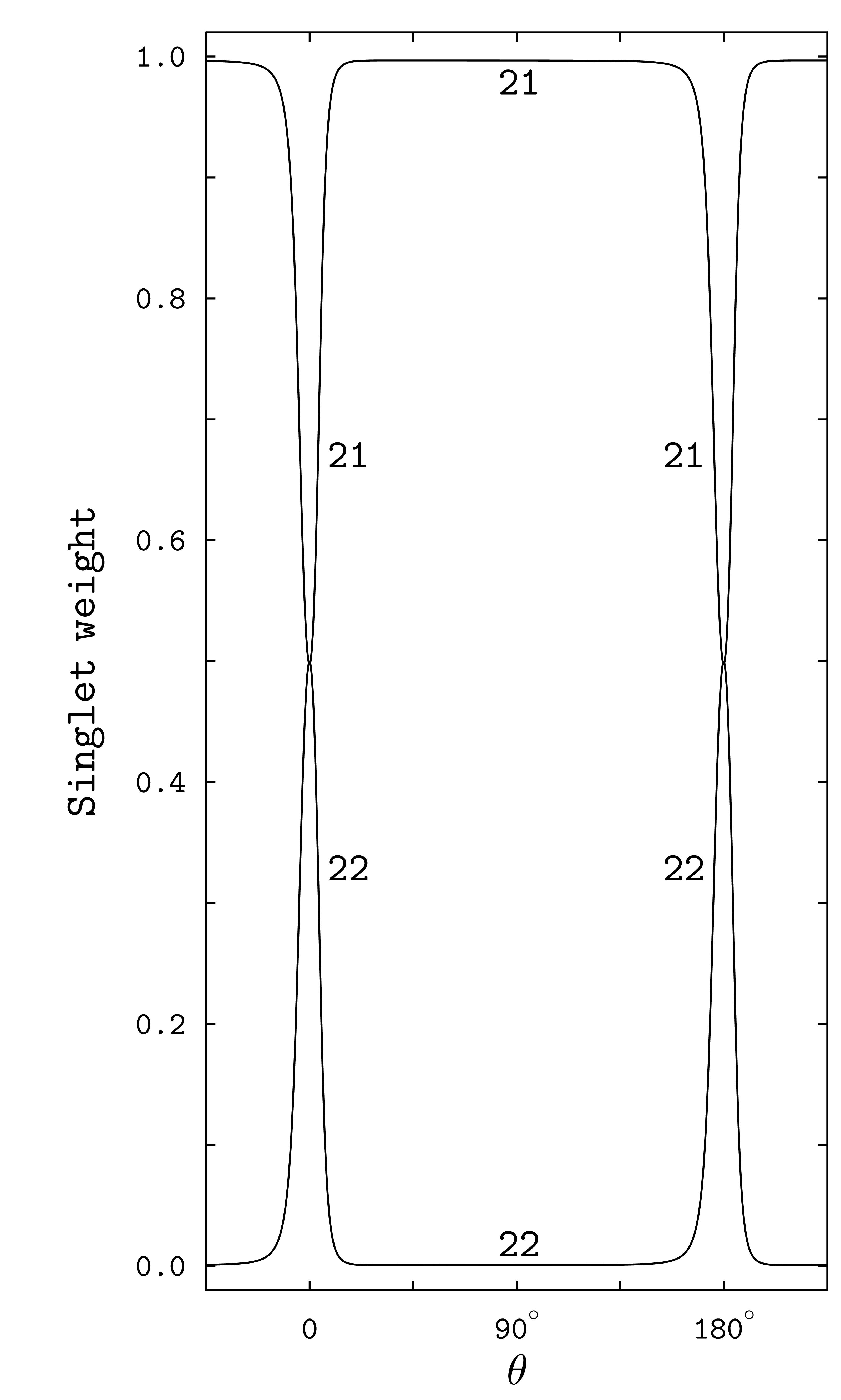}
    \caption{}
    \label{RP3_states_c}
\end{subfigure}
\caption{The effect of degeneracy of the S and T$_0$ states of the electron spin pair 
on the spin states of the RP3 model. The plots are similar to that in Figure~\ref{fig_5}, 
with the solid lines showing the energy levels and singlet weights of the RP3 states 
21 and 22.}
\label{fig_7}
\end{figure}
}

The avoided crossing of the energy levels corresponds to strong variation of the 
electron spin character of the RP3 states 21 and 22: see Figure~\ref{fig_7}(c). 
Outside the crossings, the singlet weight is almost unity for the state 21 and is almost 
zero for the state 22, whereas at the crossings it approaches the value of $0.5$ for 
both states. Given the dependence of the state energies on the magnetic direction, we 
can conclude that the spin characters of the states 21 and 22 are close to the S and T$_0$ 
types, respectively, outside the crossings, and are equally contributed by the S and T$_0$ 
types at the crossings.

\subsection{RP4 model}
\label{RP4_model}
%
% Figure 8
%
\bigskip
{\linespread{1.2}
\begin{figure}[h]
\centering
\begin{subfigure}{0.49\textwidth}
    \includegraphics[width=\textwidth]{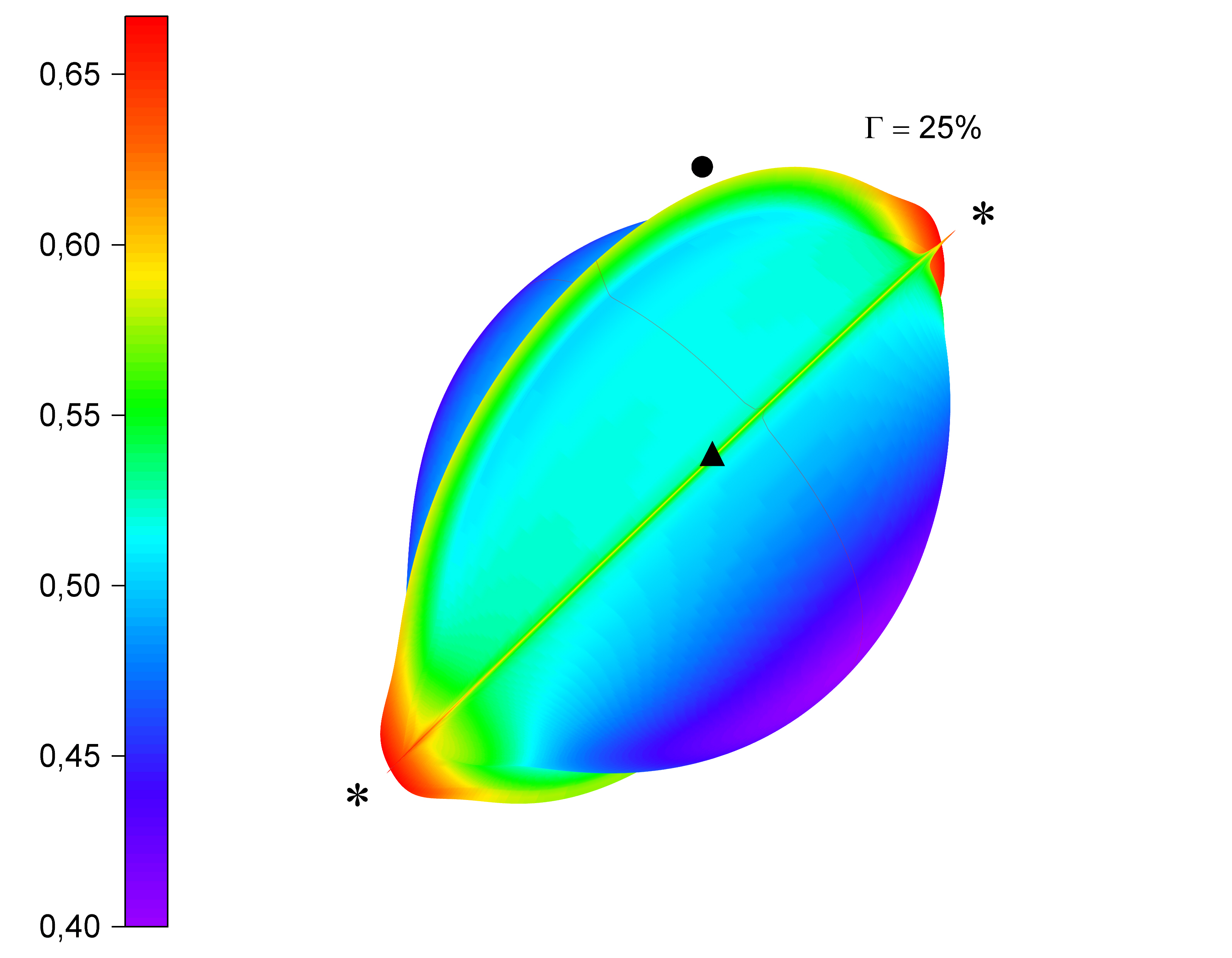}
    \caption{}
    \label{RP4a}
\end{subfigure}
\hfill
\begin{subfigure}{0.49\textwidth}
    \includegraphics[width=\textwidth]{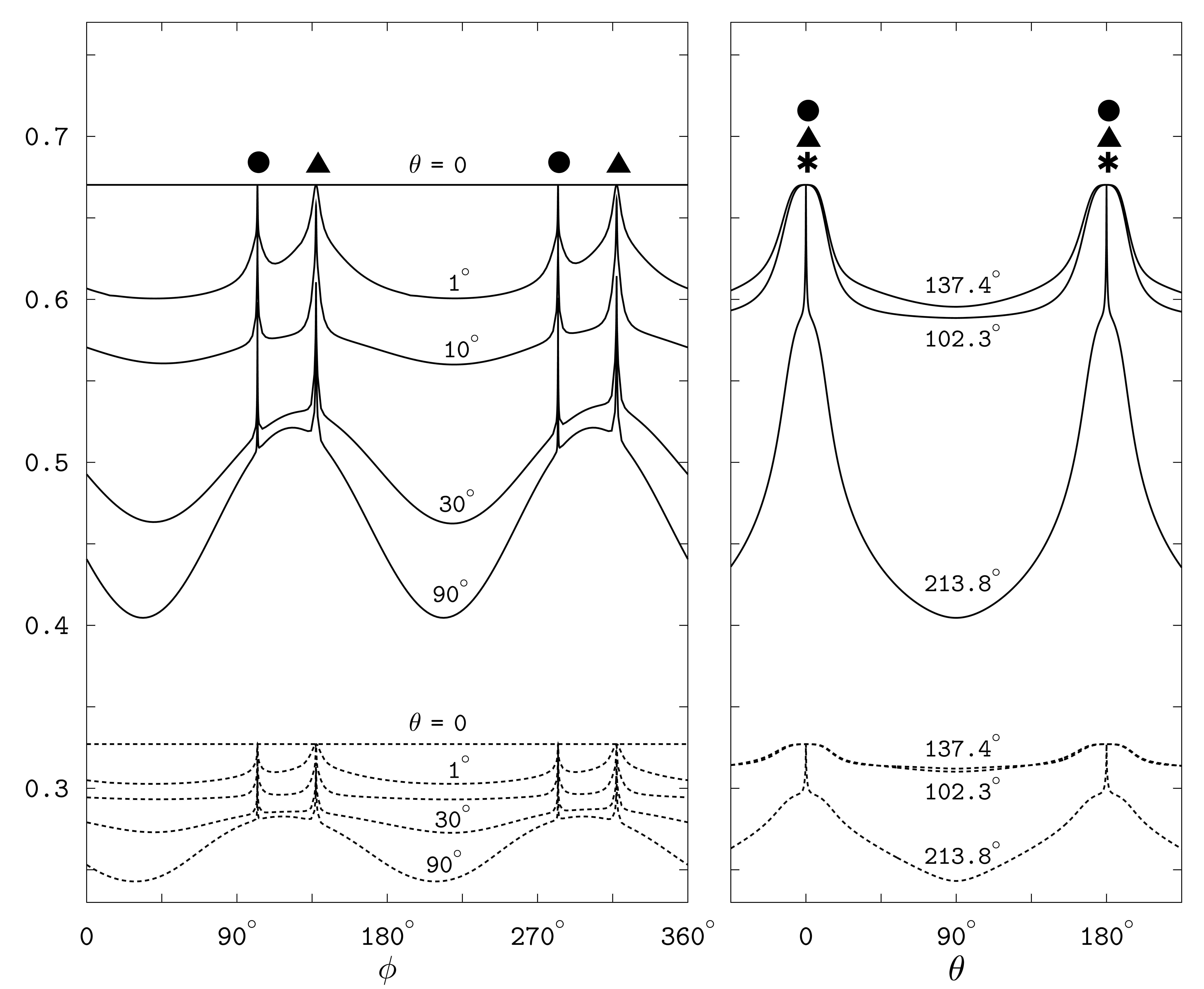}
    \caption{}
    \label{RP4b}
\end{subfigure}
\caption{The field-direction response of RP4, for the same reference coordinate system 
for the 3D polar graph as in Figures~\ref{fig_2}, \ref{fig_4}, and \ref{fig_6}, and 
the same definition of the angles $\phi$ and $\theta$ as in Figures~\ref{fig_4} and 
\ref{fig_6}, but with another selection of the 2D angular profiles. The symbols 
$\CIRCLE$ and $\blacktriangle$ indicate the spikes at the directions of the magnetic 
field orthogonal to the FAD N5 and tryptophan N1 HFC axes, respectively, and the symbol 
\raisebox{-0.45em}{\resizebox{!}{1.1em}{\bf *}} indicates the orientation of the EED axis. 
The spikes of two kinds merge together, so that the planes of the spikes in the polar 
graph cross each other, at the field directions along the EED axis, $\theta=0$ and 
$\theta=180^\circ$. 
} 
\label{fig_8}
\end{figure}
}
The model radical pair RP4 was derived from the RP3 one by rotating the radicals of the 
latter, each about the center of the density of its unpaired electron, to orient the 
HFC axis perpendicular to the EED axis, see Figure~\ref{fig_1}(4). The spin interaction 
was then modified by applying the corresponding rotations to the HFC tensors, without 
changing the EED and exchange coupling that is determined by the distance between the 
centers~\cite{Efimova-2008-03-01}. Such rotations can be performed in different 
ways, yielding a manifold of the radical pair structures with the orthogonal 
arrangements of the HFC axes relative to the EED axis, of which we have considered 
one particular case. Note that the rotations notably modify the mutual orientation of 
the radicals compared to the original orientation in RP1: the inclinations of 
the FAD and Trp HFC axes to the EED axis change from $56^\circ$ and $38^\circ$, 
respectively, in RP1 to $90^\circ$ in RP4, see also Figure~\ref{fig_1}(4). 
We introduce RP4 here as an idealized model constellation of the spin couplings 
in the radical pair, which preserves the relationship between the EED and the exchange 
spin couplings as in RP3, and additionally ensures the alignment of the 
HFC-determined special direction with the EED axis. Such a constellation results in 
a significant 25\% anisotropy of the interconvewrsion triplet yield, along with 
the appearance of spikes observed for the model RP1 without the 
inter-radical coupling. Like Figure~\ref{fig_2}, Figure~\ref{fig_8} displays the 
spikes as the rings in the 3D polar graph, and as the sharp peaks in the 2D angular 
profiles of the yield. As in RP1, the spikes appear at the magnetic field 
orientations orthogonal to the HFC axes of FAD N5 and Trp N1, and reflect thereby 
the HFC anisotropy. The FAD and Trp related spikes merge together, so that the spike 
planes in the polar graph cross each other, at the field directions parallel and 
anti-parallel to the EED axis. The same directions correspond to the maximal 
enhancement of a non-spiky part of the yield; the anisotropy of the latter part is 
similar to the entire yield anisotropy in the RP3 model, and hence is introduced 
by the EED spin coupling.
%
% Figure 9
%
\bigskip
{\linespread{1.2}
\begin{figure}[h]
\centering
\begin{subfigure}{0.32\textwidth}
    \includegraphics[width=\textwidth]{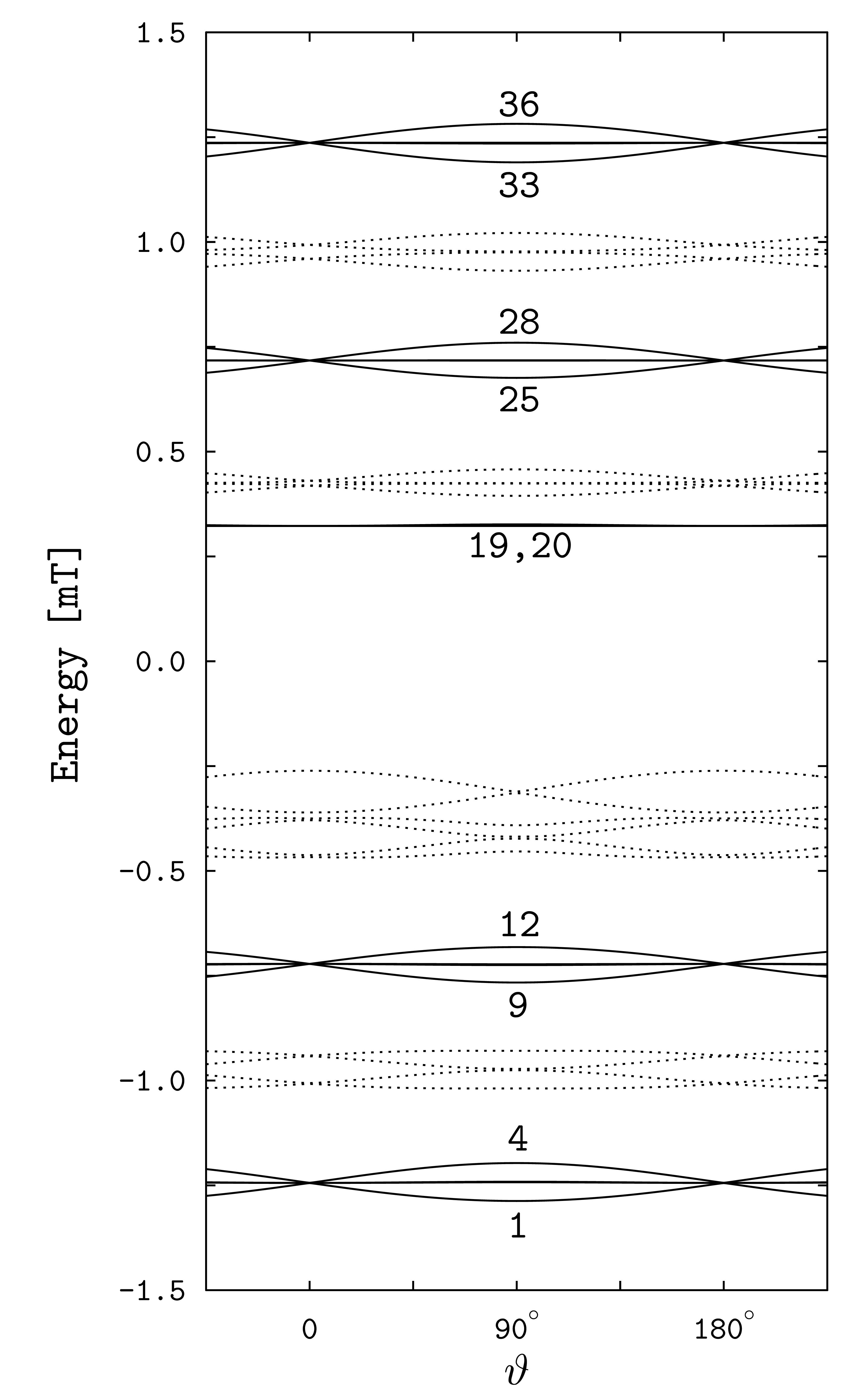}
    \caption{}
    \label{RP4_states_a}
\end{subfigure}
\hfill
\begin{subfigure}{0.32\textwidth}
    \includegraphics[width=\textwidth]{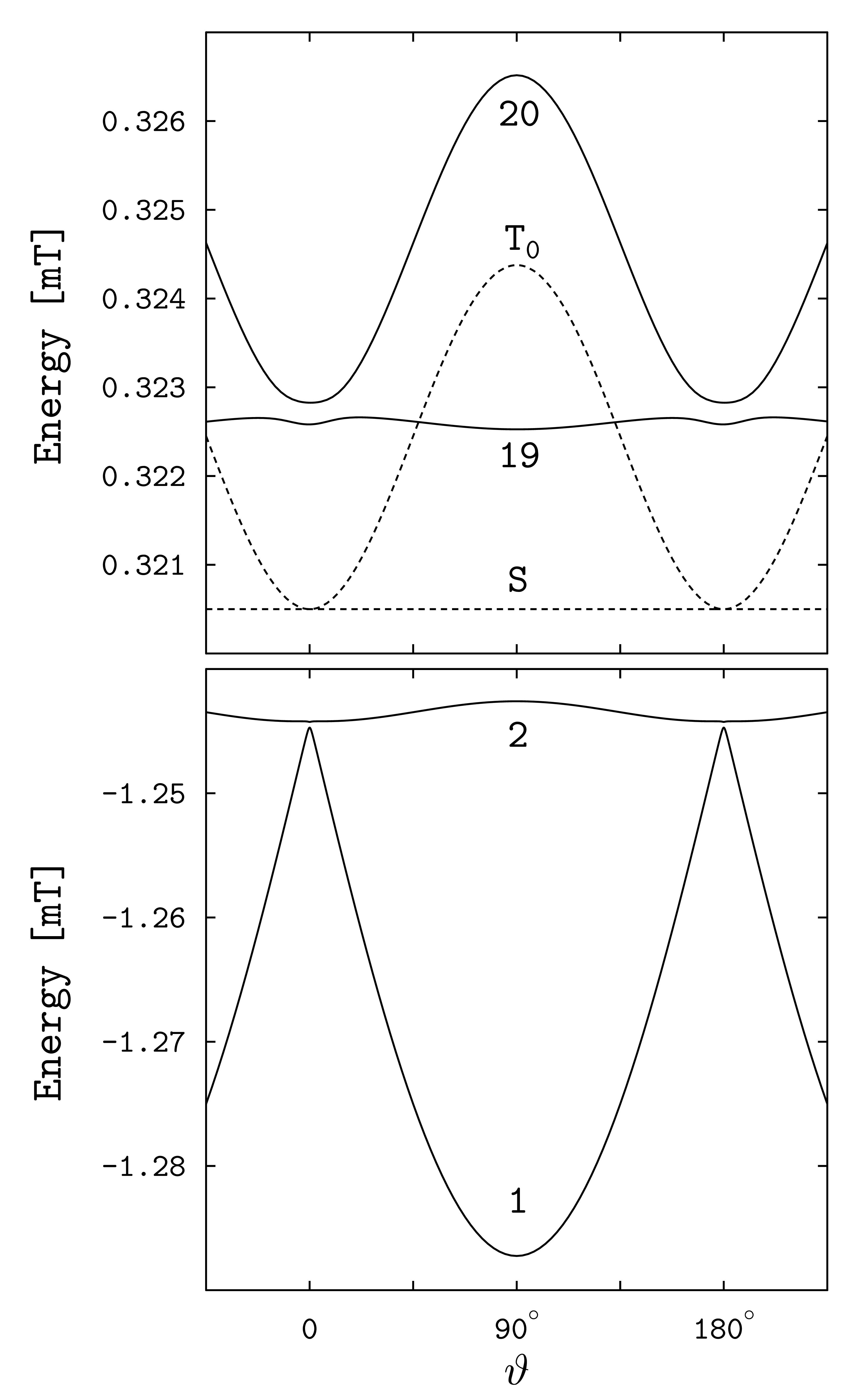}
    \caption{}
    \label{RP4_states_b}
\end{subfigure}
\hfill
\begin{subfigure}{0.32\textwidth}
    \includegraphics[width=\textwidth]{fig_9c.jpg}
    \caption{}
    \label{RP4_states_c}
\end{subfigure}
\caption{The properties of the spin states in the RP4 model. The graphs show the 
energies and singlet weights of the states as the functions of $\theta$ at 
$\phi=213.8^\circ$. 
(a) The manifold of the RP4 energy levels. The solid lines show the levels 
contributing to the anisotropy of the magnetic response of the radical pair. 
Four bunches of four levels each contribute to the spikes, and the levels 19 and 20 
contribute to the non-spiky part of the anisotropy. 
(b) Bottom plot: the avoided crossing of the levels 1 and 2 at the positions of spikes. 
Top plot: the RP4 levels 19 and 20 (solid lines) compared to the S and T$_0$ levels 
of the electron spin pair (dashed lines). 
(c) The singlet weights of the RP4 states 1 and 2 (bottom plot) and 19 and 20 (top plot).}
\label{fig_9}
\end{figure}
}

The anisotropy of magnetic response of the RP4 model stems from the properties of the 
spin states illustrated in Figure~\ref{fig_9} by the energies and singlet weights 
dependent on the angle $\theta$ between the magnetic field and the EED axis. As in 
the RP1, we find the states with the same spin-level numbers, 1 to 4, 9 to 12, 25 to 28, 
and 33 to 36, responsible for the spikes and reflecting the HFC anisotropy 
for the RP4 model. In addition, we find nearly degenerate states 19 and 20 whose properties 
result from the S--T$_0$ degeneracy of the two-electron spin states and reflect the 
EED anisotropy. This pair of RP4 states is similar to the RP3 states 21 and 22, and is 
responsible for the enhancement of the non-spiky part of the triplet yield in the 
field directions aligned with the EED axis.

\subsection{Models with expanded intra-radical coupling}
\label{RP4a_RP4b_and_other_models}

In the models considered above, the ``quantum-needle" sharpness of the compass function 
is clearly assigned to the highly axial FAD N5 and Trp N1 HFCs, whereas the particular 
RP4 arrangement of the radicals prevents this sharpness from being destroyed by 
the balanced~\cite{Efimova-2008-03-01} EED/exchange spin interaction. For the latter 
arrangement, the quantum needle effect can also develop with contributions to the 
hyperfine spin interaction that preserve the overall axial anisotropy of the 
intra-radical spin couplings. To show this, we calculated the field-direction 
response keeping the inter-radical coupling and arrangement of the radicals as in the 
RP4 model, but expanding the HFC contributions into the intra-radical spin coupling.

The expanding HFC decreases the height of the spikes and the values of the response 
anisotropy. This effect, displayed with inclusion of additional HFC contributions in 
the FAD radical, is described below for the RP4a and RP4b model radical pairs. 
In RP4a, we included the HFC contribution from the FAD nitrogen N10, 
whereas in RP4b we added the contributions from the FAD nitrogen N10 and hydrogen H8. 
Both N10 and H8 HFC tensors were computed from the same density of the flavin unpaired 
electron as the density used to compute the nitrogen N5 tensor given 
in Table~\ref{table}. The N10 HFC tensor has two small principal values 
$0.0037$ and $0.0143$~mT, and is strongly axial with respect to the principal 
direction corresponding to a large third principal value $0.7410$~mT. The latter 
direction deviates from the N5 HFC axis by a small angle of $1.63^\circ$, so that 
the N5 and N10 HFCs are nearly coaxial as noted, 
e.g., in the studies~\cite{Hiscock-2016-04-26,Deviers-2022-07-13}. 
The H8 HFC tensor has the principal values $(0.7150, 0.7254, 0.8116)$~mT close to each 
other, and hence contributes mostly to the isotropic HFC part. 

The shapes of the above described HFC tensors are shown in Figure~\ref{fig_10}, 
which compares the interconversion triplet yield of RP4a and RP4b with the yield of RP4. 
The polar graphs in the figure display two hemispheres of the yield magnitude 
depending on the magnetic field direction: the top-right hemisphere corresponds to RP4, 
whereas the bottom-left hemisphere corresponds to RP4a in the left graph, and to PR4b 
in the right graph. The equatorial plane, which divides the hemispheres, is orthogonal to 
the EED axis indicated with \raisebox{-0.45em}{\resizebox{!}{1.1em}{\bf *}}. 
Augmenting each hemisphere by its inversion with respect to the center of the dividing 
plane (i.e., the center of the coordinate frame shown in Figure~\ref{fig_2}a) 
describes the yields in the entire domain of the magnetic directions 
(i.e., for the $4\pi$ solid angle). 
%
% Figure 10
%
\bigskip
{\linespread{1.2}
\begin{figure}[h]
\centering
\includegraphics[width=\textwidth]{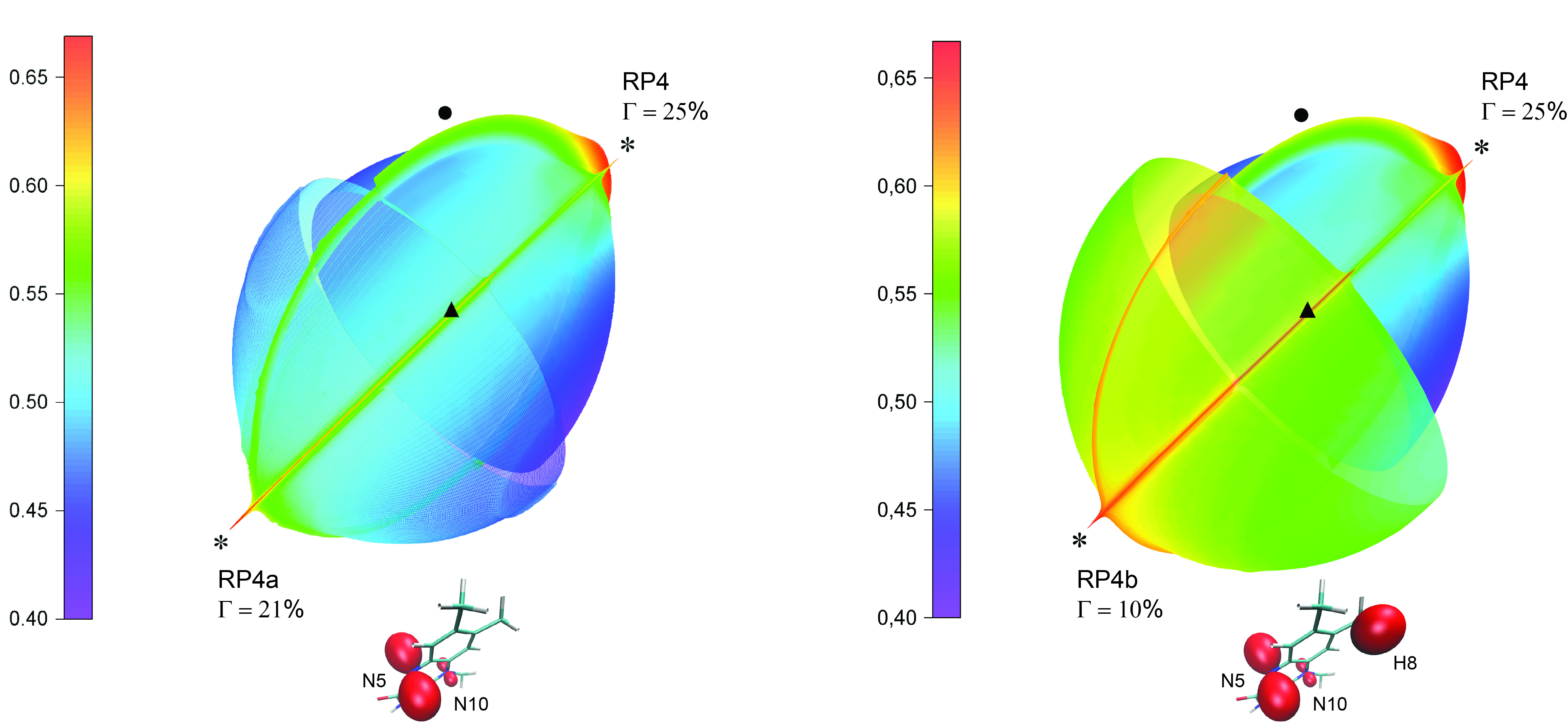}
\caption{
The comparison of the magnetic response anisotropies of RP4 and RP4a (left plot) and of 
RP4 and RP4b (right plot). The hemispheres show the magnitude of the interconversion 
triplet yield varying with the direction of the magnetic field for the compared 
radical pairs indicated in the plots along with the corresponding values of 
the anisotropy parameter $\Gamma$. Next to the hemispheres for RP4a and RP4b we display 
the structure of FAD$^{\bullet{-}}$ with the included hyperfine couplings. 
The symbols $\CIRCLE$ and $\blacktriangle$ indicate the spikes at the magnetic field 
orientations orthogonal to the FAD$^{\bullet{-}}$ N5 and Trp$^{\bullet{+}}$ N1 HFC 
axes, respectively, and the symbol \raisebox{-0.45em}{\resizebox{!}{1.1em}{\bf *}} 
indicates the orientation of the EED axis.
} 
\label{fig_10}
\end{figure}
}

The expanded FAD HFC in the RP4a and RP4B models does not destroy the spiky response 
anisotropy, and the 3D surface plots in Figure~\ref{fig_10} capture the rings formed by 
the spikes. The Trp-related spikes (labeled with $\blacktriangle$ in the plots), for 
which the axial HFC in FAD plays a supportive role~\cite{Bezchastnov-2023-01-20}, 
remain as sharp as in RP4. However, the FAD-related spikes (labeled with $\CIRCLE$) 
become slightly less sharp than in RP4, as a result of a slight difference between the 
directions of the N10 and N5 HFC axes. In addition, the isotropic part of the expanded 
HFC in FAD contributes to the non-spiky background of the radical-pair response to the 
magnetic field direction, which ``blows up'' the polar graphs of the interconversion 
triplet yield of RP4a and RP4b; for RP4b such an effect is more pronounced because of 
the contribution of a large FAD H8 HFC, see Figure~\ref{fig_10}. The increasing 
background leads to a smaller height of the spikes, both FAD- and Trp-related, 
and a decreasing value of the magnetic anisotropy $\Gamma$ from $25\%$ in RP4 
to $21\%$ in RP4a and $10\%$ in RP4b.

The FAD and Trp radicals considered in our study are similar in how they support 
the spikes via the anisotropy of the highly axial FAD N5 and Trp N1 HFC contributions, 
as well as in how they weaken the spikes by the HFC contributions that do not 
display this anisotropy pattern. For an additional illustration of this property of 
the radicals, the Supplementary Materials describe the magnetic response 
with the HFC contributions additional to those taken into account in the RP4b model. 
We also show the effect of the additional HFC contributions on the anisotropy of the 
magnetic response of the RP1 and RP3 models. 
For a better insight into the impact of expanding HFC, 
we show the results of subsequent inclusion of the contributions 
of FAD N10, FAD H8, Trp H3a, and Trp H3b (see the labels of the respective atoms 
in the radical structures in Figure S1) in the basic HFC determined by the FAD N5 
and Trp N1 contributions.

As expected, the HFC contributions in the Trp radical progressively 
enhance the low-anistropy part of the magnetic response and weaken the sharpness of 
the spikes, as the FAD H8 contribution did, 
see Figures~S2--S6 of the Supplementary Materials. The diminished sharpness is 
accompanied by the appearance of a multi-spike structure, which is most prominent 
at the field directions around the directions of the FAD- and Trp-related spikes in 
the RP1 and RP4 models. This complicates producing the accurate 3D polar graphs by 
using the approach based on the fixed azimuthal angle values for the spike directions 
in the coordinate frame with the $xy$-plane defined by the anisotropy axes of the 
basic HFC (cf. Section~\ref{RP1_model}). The Supplementary Materials present, 
rather than the 3D polar plots, the 2D angular profiles of the magnetic response, 
like those shown in Figures~\ref{fig_2}(b), \ref{fig_4}(b), \ref{fig_6}(b), 
and \ref{fig_8}(b). Given the increasing number of quantum spin states, we did 
not analyze, as we did for the basic models, the spin-state properties of the models 
with expanded HFC. We also did not quantify the magnetic anisotropy for the models 
with the HFC expanded beyond that of RP4b, since obtaining accurate $\Gamma$ values 
would require careful probing of the field directions in the range of the angles within 
an entire hemisphere.

The expanding HFC promptly suppresses the magnetic anisotropy indicating the EED axis 
in the RP3 model, leading to the disappearance of the enhancement of the interconversion 
triplet yield at the magnetic field directions along this axis: see the $\theta$-profiles 
for the RP3a--RP3d models in Figures S3--S6. The spike-related anisotropy of the 
magnetic response in the RP1 and RP4 models turns out to be less sensitive to the 
expansion of the HFC contributions, so that Figures S3--S6 display the spike 
structures, though of a gradually reduced contrast, for the RP1a--RP1d and RP4a--RP4d 
series of the models. Thus, the inter-radical spin interaction and the arrangement 
of the radicals introduced in the RP4 model remain optimal in supporting the spikes 
at expanding HFC. 

\section{Discussion and conclusions}
\label{Conclusions}

In this work we have considered several models of a putative flavin-tryptophan radical 
pair of the cryptochrome chemical compass. The focus of our study was on the sensitivity 
of the spin dynamics in the radical pair to the direction of the geomagnetic field, which 
is the property demanded by the radical-pair based magnetic compass. Since previous 
studies, discussed in Section~\ref{Intro}, have demonstrated that the sensitivity 
strongly depends on the properties and interplay of the hyperfine interaction in the 
radicals and the inter-radical electron spin exchange and dipolar interaction, 
we considered a set of arrangements of these interactions, and simplified the models 
so that we could unambiguously assign the origin of a sharp magnetic response to 
specific spin interactions. We were particularly interested in finding a 
response which could serve as a compass needle for the cryptochrome magnetoreceptor. 
The models we have studied allowed an accurate 3D visualization of such sharp features 
if they develop, as well as identifying the radical-pair spin states that 
contribute to the magnetic response anisotropy. 
The key features described in this respect are the avoided 
crossings of the spin-energy levels linked to the modulation of the singlet weights by 
the magnetic field direction. These properties additionally emphasize the 
quantum nature~\cite{Fay-2019-12-13,Hiscock-2016-04-26} of the cryptochrome 
radical-pair compass.

Consistent with the original theoretical finding of the 
``quantum needle''~\cite{Hiscock-2016-04-26}, our simulations show that the highly 
axial HFCs introduced by the nitrogen atoms N5 in FAD and N1 in Trp result in the 
sharp structures in the response of the radical pair to the magnetic field direction. 
The 3D polar plots display these structures as plane rings marking the field 
directions orthogonal to the N5 and N1 HFC axes, and suggest assigning the compass 
needle to the intersection of the rings. Such a sharp magnetic response is clearly 
visible because the two axial HFCs predominantly (or exclusively, as in the idealized RP1 
model) determine the spin coupling in the radical pair.

The inclusion of the EED spin coupling corresponding to the separation of flavin and 
tryptophan in the putative cryptochrome radical pair causes the anisotropy of the 
field-direction response to almost vanish, and completely suppresses the sharp structures, 
in agreement, for example, with the results~\cite{Babcock-2020-04-02}. Such a 
weak sensitivity to the magnetic field direction is displayed by the radical pair with 
only three spin couplings, all of them axial: the intra-radical N5 and N1 HFCs and the 
inter-radical EED spin coupling included in the RP2 model. A weak sensitivity of the 
radical pair with even more spin interactions is therefore to be expected, even if these 
are highly axial. However, adding the electron exchange spin coupling, which is tuned so 
that the total inter-radical coupling supports the S--T$_0$ degeneracy of the two-electron 
spin states~\cite{Efimova-2008-03-01}, introduces a magnetic response anisotropy 
defined by the EED axis. This anisotropy is less sharp than the ``quantum needle'' one, 
but is still pronounced enough for the compass needle in the RP3 model to be assigned to 
the direction defined by the centers of the unpaired electron spin density in the radicals.

The particular finding of the present study results from an additional tuning of the 
mutual arrangement of the axial hyperfine and the EED couplings. Orienting the HFC axes 
orthogonal to the EED axis, with the EED and exchange couplings partially compensating 
each other to support the S--T$_0$ degeneracy, restores the spikes. In the 3D polar graph, 
the respective sharp structures cross each other in the directions of the EED axis. 
The RP4 model optimized in this way displays the most pronounced and sharpest anisotropy 
of the magnetic response; here both the hyperfine and EED couplings establish the direction 
in the radical-pair structure, namely the EED axis, which can be used for an 
unambiguous orientation of the compass along the magnetic field lines.

The sharpness of the radical-pair magnetic response considered in our study is based on 
certain properties of the intra- and inter-radical spin interactions and their 
mutual arrangement. Sharp ``quantum needle'' structures in the compass output require 
the hyperfine interaction, which defines one distinct HFC axis in each of the 
radicals and whose anisotropic part prevails over the isotropic one. 
This was the case for the RP1 and RP4 models, where the HFC in each radical was modeled 
by a single highly axial term. The sharp magnetic response of the extended RP4a 
and RP4b models was based on additional HFC contributions that did not disrupt 
the existence of the distinct HFC anisotropy axis for the FAD radical. 
A non-negligible EED/exchange spin interaction was found to not destroy but rather to 
enhance the ``quantum needle'' magnetic response, provided by 
the balance~\cite{Efimova-2008-03-01} of the EED and exchange interaction and the 
orthogonal orientation of the intra-radical anisotropy axes with respect to the EED axis. 
In the underlying physics, the spin interactions with the above properties are responsible 
for the quantum spin states contributing to the ``signal'': in the presented 3D plots the 
signal has the form of ``quantum needle'' rings crossing each other at the magnetic 
field direction along the EED axis. The interactions not satisfying the 
aforementioned constraints play a major role in the properties of the quantum states 
that contribute to the background ``noise''. Increasing the complexity of the 
spin interaction in the radical pair enhances the noise and weakens the signal, as 
demonstrated by the models with expanding HFC contributions in the Trp radical.

Notably, the arrangement of the anisotropic spin couplings described in our study of the 
basic radical-pair models results in the signal that delivers ``strongly anisotropic 
but relatively simple directional information''~\cite{Rodgers-2009-01-13} suggested 
earlier as the information favored for the magnetic sensing. 
However, theoretical simulations aimed at increasing the complexity of the spin 
interaction still leave open the question of whether the signal can outperform the noise 
for the putative magnetically sensitive cryptochrome radical pairs. In this regard, 
the conditions leading to sharp magnetic sensitivity in the models we studied may be 
important for further insight into the basis of cryptochrome magnetoreception.

\section*{Data availability}

The data underlying this study are available in the article and Supplementary 
Materials.

\section*{Acknowledgements}

We are grateful to Ilme Schlichting for the constant encouragement and to John Wray for 
the careful reading of the manuscript, stimulating discussions and valuable comments. 
T.D. acknowledges that her contribution in the study was conducted under the 
state assignment of Lomonosov Moscow State University, project Nr. 121031300176-3.

\section*{Contributions}

V.B. and T.D. designed the general concept of the study and selected the radical-pair 
models. V.B. performed numerical simulations, analytical derivations, and designed 
graphical 3D representation of the radical-pair response to the magnetic field 
direction. Both authors equally contributed to analysis of the results and preparing 
the manuscript.

\section*{Competing interests}

The authors declare no competing interests.

\end{document}